\documentclass[11pt,a4paper]{article}
\usepackage{jheppub,xcolor}
\usepackage[utf8]{inputenc}
\graphicspath{{./figs/}}

\usepackage{amssymb}
\usepackage{amsmath}
\usepackage{dcolumn}	
\usepackage{bm}			
\usepackage{bbm} 		
\usepackage{multirow}
\usepackage{slashed}
\usepackage{blkarray}
\usepackage{booktabs}
\usepackage{makecell}
\usepackage{longtable}
\usepackage{placeins}
\usepackage{graphicx}
\usepackage{float}
\usepackage{multirow}
\usepackage{verbatim}
\usepackage{hyperref}
\usepackage{subcaption}
\usepackage{lscape} 
\usepackage{tablefootnote} 
\usepackage[font=small,labelfont=bf]{caption}
\usepackage[compat=1.1.0]{tikz-feynman}
\usetikzlibrary{positioning}
\usepackage{siunitx}

\usepackage[normalem]{ulem}

\usepackage{lineno}

\usepackage[compat=1.1.0]{tikz-feynman}
\pgfmathsetmacro\sizedot{1.1}
\pgfmathsetmacro\sizesqdot{1.6}
\pgfmathsetmacro\sizeemdot{2.}

\pgfmathsetmacro\lwL{2.1}
\pgfmathsetmacro\lwR{.4}

\usepackage{xparse}
\ExplSyntaxOn
\NewDocumentCommand{\longdash}{ O{2} }
{
	--\prg_replicate:nn { #1 - 1 } { \negthinspace -- }
}
\ExplSyntaxOff

\usepackage[compat=1.1.0]{tikz-feynman}
\pgfmathsetmacro\sizedot{1.1}
\pgfmathsetmacro\sizesqdot{1.6}
\pgfmathsetmacro\sizeemdot{2.}

\pgfmathsetmacro\lwL{2.1}
\pgfmathsetmacro\lwR{.4}

\definecolor{nicered}{rgb}{0.6,0.1,0.1}
\definecolor{nicegreen}{rgb}{0.1,0.5,0.1}
\definecolor{mediumcandyapplered}{rgb}{0.99, 0.12, 0.07}
\definecolor{red}{rgb}{1.0, 0, 0}
\hypersetup{colorlinks,citecolor= nicegreen,linkcolor= nicered}

\newcommand{\coeff}[2]{ \mathcal{C}_{#1} ^{#2} }
\newcommand{\op}[2]{ \mathcal{O}_{#1} ^{#2} }

\newcommand{\PR} {\mathbb{P}_{\rm R}}
\newcommand{\PL} {\mathbb{P}_{\rm L}}
\newcommand{\order}[1] {\mathcal{O }\left( #1 \right)}
\newcommand{\yuk}[1]{{Y}_{#1}}
\newcommand{\hc}{\text{H.c.}}

\newcommand{\lround}{\left(}
\newcommand{\rround}{\right)}
\newcommand{\lsquare}{\left[}
\newcommand{\rsquare}{\right]}

\makeatletter
\newcommand{\overleftrightsmallarrow}{\mathpalette{\overarrowsmall@\leftrightarrowfill@}}
\newcommand{\overrightsmallarrow}{\mathpalette{\overarrowsmall@\rightarrowfill@}}
\newcommand{\overleftsmallarrow}{\mathpalette{\overarrowsmall@\leftarrowfill@}}
\newcommand{\overarrowsmall@}[3]{%
	\vbox{%
		\ialign{%
			##\crcr
			#1{\smaller@style{#2}}\crcr
			\noalign{\nointerlineskip}%
			$\m@th\hfil#2#3\hfil$\crcr
		}%
	}%
}
\def\smaller@style#1{%
	\ifx#1\displaystyle\scriptstyle\else
	\ifx#1\textstyle\scriptstyle\else
	\scriptscriptstyle
	\fi
	\fi
}
\makeatother
\newcommand{\lrarrow}[1]{\overleftrightsmallarrow{#1}}

\title{
Complete two-loop Yukawa-induced running of the Higgs-gluon coupling in SMEFT
}

\author[a]
{Stefano Di Noi }
\author[b,c]
{, Barbara Anna Erdelyi }
\author[b,c]
{and Ramona Gr\"{o}ber }

\affiliation[a]{Institute for Theoretical Physics, Karlsruhe Institute of Technology (KIT),\\
Wolfgang-Gaede Stra\ss{}e 1, Karlsruhe D-76131, Germany}
\affiliation[b]{Dipartimento di Fisica e Astronomia "G. Galilei", Università degli Studi di Padova,\\
Via F. Marzolo 8, 35131 Padova, Italy}
\affiliation[c]{Istituto Nazionale di Fisica Nucleare, Sezione di Padova, \\
Via F. Marzolo 8, 35131 Padova, Italy}

\emailAdd{stefano.dinoi@kit.edu, barbaraanna.erdelyi@phd.unipd.it, ramona.groeber@pd.infn.it}
\preprint{KA-TP-31-2025,P3H-25-076, COMETA-2025-47}

\date{\today}

\abstract{
 We compute the two-loop renormalisation group equation for the effective Higgs to gluon coupling in Standard Model Effective Field Theory. Concretely, we present the contributions generated by the operators belonging to class 3 and 7 in the Warsaw basis, completing the two-loop renormalization program of the Higgs-gluon coupling proportional to the top Yukawa coupling for potentially tree-level generated operators.
We investigate the phenomenological impact of the contributions in fits to Higgs data both in a bottom-up approach and a top-down approach in terms of UV models with vector-like quarks.
}

\begin{document}

\maketitle

%




\section{Introduction}
There are several reasons to believe that the Standard Model (SM) of particle physics requires an extension. Despite this, no direct evidence of new physics has been observed so far, motivating the use of the Effective Field Theory (EFT) framework to describe potential beyond-the-SM (BSM) effects in the most model-independent way possible. 
This approach is valid under the assumption that the scale of new physics lies well above the electroweak scale. 
Assuming that the Higgs boson transforms as in the SM, namely as part of an $\mathrm{SU(2)_L}$ doublet, writing down all possible higher-dimensional operators consistent with the symmetries of the SM leads to the so-called Standard Model EFT (SMEFT).
Concerning collider physics, the most relevant contributions are expected to arise from dimension-six operators, for which a non-redundant basis was first established in Ref.~\cite{dim6smeft}, the so-called Warsaw basis.

Despite not being the only possible EFT extension of the SM,\footnote{A different philosophy is embodied by the Higgs EFT \cite{Feruglio:1992wf, Longhitano:1980iz, Grinstein:2007iv, Buchalla:2013rka,Brivio:2013pma,Brivio:2014pfa,Alonso:2014wta} (HEFT or electroweak chiral Lagrangian), where the Goldstone bosons and
the Higgs boson do not transform in the same multiplet, but the physical Higgs boson transforms as a singlet. 
} the SMEFT has become a popular framework for global analyses \cite{Bartocci:2023nvp,deBlas:2025xhe,Celada:2024mcf,Bellafronte:2025jbk, ATLAS:2024lyh}, providing a self-consistent way to correlate various experimental measurements across different energy regimes.

The connection between the Wilson Coefficients (WCs) at different energy scales is encoded in the renormalisation group equations (RGEs). The complete one-loop RGEs of dimension-six operators have been computed in \cite{rge1, rge2, rge3} and implemented for systematic inclusion into phenomenological analyses in several computer codes \cite{Lyonnet:2013dna, Celis:2017hod, Aebischer:2018bkb, Fuentes-Martin:2020zaz, DiNoi:2022ejg}. 
Even though the full set of two-loop RGEs for the dimension-six SMEFT has not been computed yet, many partial results have been presented in the last years 
and progress towards the two-loop SMEFT RGEs has been reported  \cite{Bern:2020ikv, Jenkins:2023rtg, DiNoi:2023ygk, Jenkins:2023bls, Fuentes-Martin:2023ljp, DiNoi:2024ajj, Born:2024mgz, Duhr:2025zqw, Haisch:2025lvd, Assi:2025fsm,Haisch:2024wnw,DiNoi:2025arz,Duhr:2025yor,Banik:2025wpi, Haisch:2025vqj}. 

The RGEs are thus crucial for global analyses that combine observables measured at different energy scales \cite{deBlas:2015aea,terHoeve:2025gey, Bartocci:2024fmm}, exploiting fully the variety of available data from different experiments.
Moreover, with the increasing precision of collider measurements, RGE effects can also become sizeable in processes where the renormalisation scale varies dynamically over a wide range of energies. In this context, including the RGE running of WCs is essential, as explicitly demonstrated in \cite{Aoude:2022aro, Maltoni:2024dpn, Grazzini:2018eyk, Battaglia:2021nys, Heinrich:2024rtg}. These works focus on the running proportional to the strong coupling constant $\alpha_s$, which is expected to give the dominant contributions. Moreover, restricting to these effects allows to solve the RGEs analytically, greatly improving the time efficiency of the computer programs employed for phenomenological studies.

However, also the running effects generated by the top-quark Yukawa coupling $\yuk{t}$ can produce significant effects \cite{DiNoi:2023onw,DiNoi:2024ajj}.
Despite the strength of the interaction being na\"ively smaller than in the strong case, WCs entering the RGEs via Yukawa-induced contributions are generally less constrained \cite{Celada:2024mcf}, potentially enhancing their impact. 

In this work we complete the effort started in Refs.~\cite{DiNoi:2023ygk,DiNoi:2024ajj} to determine the two-loop $\order{g_s^2 \yuk{t}^2}$ RGE of the Higgs-gluon coupling, which enters the dominant gluon-fusion production channel for Higgs production. Within the SMEFT framework, this coupling is parametrized by the operator
$\op{HG}{} = (H^\dagger H) G_{\mu \nu}^A G^{A,\mu\nu}$
where $H$ denotes the Higgs doublet and $G_{\mu\nu}^A$ the gluon field strength. We stress that this interaction generates a tree-level coupling between the Higgs bosons and the gluons, which instead is loop induced in the SM. 
In weakly interacting and renormalizable UV models, the corresponding WC $\coeff{HG}{}$ is loop-generated \cite{Arzt:1994gp,Buchalla:2022vjp}. Moreover, at one-loop level, the RGE of $\coeff{HG}{}$ involves only WCs of operators that are likewise loop-generated under the same assumptions. 
Consequently, $\op{HG}{}$ does not undergo one-loop renormalisation, even when considering dimension-eight effects \cite{Grojean:2024tcw}.
It follows that, to ensure consistency in the perturbative expansion in the loop order, the two-loop running effects proportional to tree-level generated operators must be considered when including RGE effects for Higgs production.  
Contributions belonging to this category include the ones from four-quark operators \cite{DiNoi:2023ygk, Haisch:2025vqj} and Yukawa-like operators \cite{DiNoi:2024ajj}.

In this paper, we present the contributions from the operators with the schematic structure $D^2 H^4$ and $D\, H^2 \,\psi^2$. Such operators belong to, respectively, class 3 and 7 in the classification of the Warsaw basis notation~\cite{dim6smeft} and are the last missing contributions for the complete two-loop RGE proportional to the third family Yukawa couplings $\yuk{t}$ and $\yuk{b}$ involving operators that are potentially tree-level generated. Conversely, we do not consider chromomagnetic operators as they can only be generated at loop-level in weakly-interacting theories \cite{Arzt:1994gp,Buchalla:2022vjp} and are expected to give a contribution at $\mathcal{O}(g_s^2 Y_t)$ that would be a three-loop effect. 

We investigate the phenomenological impact of the new contributions to the RGE running by performing a fit to Higgs data both in the bottom-up and top-down approach, using simplified models with heavy Vector-Like Quarks (VLQs).
Such states naturally arise in a variety of frameworks proposed to address open issues of the Standard Model, such as the hierarchy problem, with Composite Higgs Models providing a prominent example \cite{Agashe:2004rs, DeSimone:2012fs, Gillioz:2012se}.
In these models, operators 
belonging to class 7 are generated at tree level, which lead to a modification of the couplings of quarks to massive vector, eventually in association with a Higgs boson. 
In particular, the couplings of top quarks to $Z$ bosons are weakly constrained so far \cite{terHoeve:2025gey, Ellis:2020unq}, as a direct constraint comes from $t\bar{t}Z$ production, single top production \cite{Brivio:2019ius} or $Zh$ production \cite{Englert:2016hvy}, whereas other probes need to rely on RGE effects. Other operators that couple to the third generation $\mathrm{SU(2)}_{\mathrm{L}}$ doublet can also be constrained by electroweak precision data.
Conversely, class 3 operators can be constrained by electroweak precision data and by measurements of the Higgs coupling to massive vector bosons, currently available at few percent level \cite{ATLAS:2022vkf, CMS:2022dwd}. 
\par

The paper is structured as follows. In Sec.~\ref{sec:notation}  we present our notation. In Sec.~\ref{sec:computation} we describe our computation and present the result for the RGE of the Higgs-gluon coupling in the SMEFT. In Sec.~\ref{sec:pheno} we study the phenomenological implications of the inclusion of two-loop running effects in single Higgs physics. Finally, in Sec.~\ref{sec:conclusions} we present our summary.

\section{Notation} \label{sec:notation}
The SMEFT extends the SM by a tower of gauge and Lorentz invariant higher-dimensional operators as a series in inverse powers of a high-energy new physics scale $\Lambda$. In this work, we neglect lepton and baryon number violation and focus exclusively on the leading dimension-six terms in the expansion:
\begin{equation}
\mathcal{L} = \mathcal{L}_{\text{SM}} 
+ \sum_{\mathcal{D}_i=6} {\coeff{i}{}} \op{i}{} .
\end{equation}
The operators are denoted by $\op{i}{}$ and the associated WCs by $\coeff{i}{}$. We note that the WCs in the previous formula have dimension -2 in units of energy.
The operators in the previous expressions are invariant under the unbroken SM group $\mathcal{G}_{\mathrm{SM}}=\mathrm{SU(3)_C} \times \mathrm{SU(2)_L} \times \mathrm{U(1)_Y}$. 

We follow the Warsaw basis of Ref.~\cite{dim6smeft} for the dimension-six operators and Ref.~\cite{rge1} for the definitions of the SM Lagrangian. 

We consider only the third family of quarks, leading to a minimal modification of the notation with respect to the original publication. In particular, we use $t_R,b_R$ to denote the right-handed component and $Q_L$ to denote the left-handed component of the third family of quarks. 
The relevant class 7 operators are defined as:
\begin{align}
   \op{H Q}{(1)} & = \lround iH^\dagger \lrarrow{D}_\mu H\rround \lround \bar{Q}_{L} \gamma^\mu Q_L \rround\,, \quad \op{H Q}{(3)}  = \lround iH^\dagger \lrarrow{D}^I_\mu H\rround \lround \bar{Q}_{L} \gamma^\mu\tau^I Q_{L} \rround\,, \label{eq:psi2phi2Dops_phiqL}\\
   \op{H t}{}&= \lround i \, H^\dagger \lrarrow{D}_\mu H \rround \lround \bar{t}_{R} \gamma^\mu t_{R} \rround \,, \quad\op{H b}{} = \lround i \, H^\dagger \lrarrow{D}_\mu H \rround \lround \bar{b}_{R} \gamma^\mu b_R \rround\,,\label{eq:psi2phi2Dops_phiqR}\\
   &\qquad \qquad  \op{H tb}{} = \lround i \, \tilde{H}^\dagger D_\mu H \rround \lround \bar{t}_{R} \gamma^\mu b_{R} \rround\, , \label{eq:psi2phi2Dops_phiud}
\end{align}
where $\tilde{H}_i=\epsilon_{ij}H^{\dagger, j}$ (with $\epsilon_{12}=+1$) and 
\begin{align}
\lround i \, H^\dagger \lrarrow{D}_\mu H \rround  &= i H^\dagger (D_\mu H ) - i (D_\mu H )^\dagger H  , \\ 
\lround i \, H^\dagger \lrarrow{D}_\mu ^I H \rround  &= i H^\dagger \tau^I (D_\mu H ) - i (D_\mu H )^\dagger \tau^I H. 
\end{align}

 The operators in class 3 are defined as 
\begin{equation} \label{eq:phi4d2}
\op{H\Box}{} = \left(H^\dagger H\right) \Box \left(H^\dagger H\right), \quad \op{HD}{} = \left(H^\dagger D_\mu H \right) \left(H^\dagger D^\mu H \right)^*.
\end{equation}
These two operators in the broken phase modify the kinetic term of the Higgs boson, thus requiring a field redefinition to reobtain the canonical normalisation of the Higgs kinetic term. In the unitary gauge, the Higgs doublet $H$ is related to the physical Higgs field $h$ via
\begin{equation}
    H = \frac{1}{\sqrt{2}}\begin{pmatrix}
        0 \\ v + h \left( 1 + v^2 \,\coeff{H,\, \text{kin}}{} \right)
    \end{pmatrix}\,, \quad \text{where} \,\,\coeff{H,\, \text{kin}}{} = \coeff{H\Box}{} - \frac{1}{4}\coeff{HD}{}\,.  \label{eq:Hkin_definition}
\end{equation}

Finally, we recall that the Higgs-gluon coupling in the SMEFT is parametrized by the operator
\begin{equation}\label{eq:OHG}
\op{HG}{} = \left( H^\dagger H \right) \left(G^A_{\mu\nu} G^{A\, \mu\nu} \right)\,.
\end{equation}
This operator belongs to class 4 in the Warsaw basis. Other class 4 operators that will appear in the phenomenological study performed in this work are defined below
\begin{gather}
    \op{HW}{} = \lround H^\dagger H\rround \lround W^I_{\mu\nu} W^{I\,\mu\nu} \rround\,, \quad \op{HB}{} = \lround H^\dagger H\rround \lround B_{\mu\nu} B^{\mu\nu} \rround\,,\label{eq:def_class4_HWHB}\\
    \op{HWB}{} = \lround H^\dagger \tau^I H \rround \lround W_{\mu\nu}^I B^{\mu\nu}\rround\,.\label{eq:def_class4_HWB}
\end{gather}
Furthermore, the following class 5 operators will be considered, often referred to as Yukawa-like operators:
\begin{equation}
    \op{tH}{} = \lround H^\dagger H\rround \lround \bar{Q}_L \tilde{H} t_R \rround\,, \quad \op{bH}{} = \lround H^\dagger H\rround \lround \bar{Q}_L H b_R \rround\,. \label{eq:def_class5}
\end{equation}
We remind the reader that the SM Yukawa coupling is not defined in direct analogy to the Yukawa-like operator.

All the previous operators are Hermitian with the exclusion of $\op{Htb}{}$, $\op{tH}{}$ and $\op{bH}{}$, which enforce the addition of their complex conjugate  to the Lagrangian.
We use $T^A$ for the $\mathrm{SU(3)_C}$ generators and $\tau^I$ for the Pauli matrices. We define the gluon field-strength tensor as  
$ G_{\mu\nu}^A = \partial_\mu G_\nu^A - \partial_\nu G_\mu^A - g_s f^{ABC} G_\mu^B G_\nu^C$. 

\section{Computation and result} \label{sec:computation}

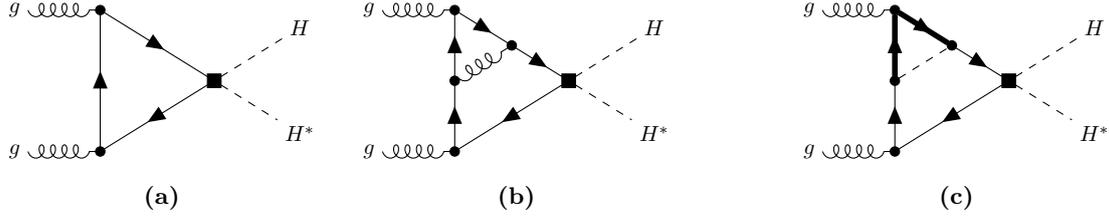
\begin{figure}
\centering
 \begin{subfigure}[t]{0.3\textwidth}\centering
        \begin{tikzpicture}
            \begin{feynman}[scale=0.75, transform shape]
              \vertex (vert) at (0,0) [square dot, scale=\sizesqdot, black] {};
               \vertex (phi) at (1.5, .9375) {\(H\)};
               \vertex (phistar) at (1.5, -.9375) {\(H^*\)};
               \vertex (Gamu) at (-3.5, 1.25) {\(g \)};
               \vertex (Gbnu) at (-3.5, -1.25) {\( g \)};
               \vertex (v1) [dot, scale = \sizedot ] at (-2, 1.25) {};
               \vertex (v2) [dot, scale = \sizedot ] at (-2, -1.25)  {};
                \diagram*{
                    (phi) -- [scalar] (vert) -- [scalar] (phistar);
                    (Gamu) -- [gluon] (v1);
                    (Gbnu) -- [gluon] (v2);
                    (v1) -- [fermion, line width=\lwR] (vert) -- [fermion,line width=\lwR] (v2) -- [fermion,line width=\lwR]  (v1); 
                };
            \end{feynman}    
        \end{tikzpicture}
    \caption{}\label{fig:discarded1L}
    \end{subfigure}
 \begin{subfigure}[t]{0.3\textwidth}\centering
        \begin{tikzpicture}
            \begin{feynman}[scale=0.75, transform shape]
              \vertex (vert) at (0,0) [square dot, scale=\sizesqdot, black] {};
               \vertex (phi) at (1.5, .9375) {\(H\)};
               \vertex (phistar) at (1.5, -.9375) {\(H^*\)};
               \vertex (Gamu) at (-3.5, 1.25) {\(g \)};
               \vertex (Gbnu) at (-3.5, -1.25) {\( g \)};
               \vertex (v1) [dot, scale = \sizedot ] at (-2, 1.25) {};
               \vertex (v2) [dot, scale = \sizedot ] at (-2, -1.25)  {};
               \vertex (Yuk1) [dot, scale = \sizedot ] at (-2, 0.)  {};
               \vertex (Yuk2) [dot, scale = \sizedot ] at (-1., .625) {};
                \diagram*{
                    (phi) -- [scalar] (vert) -- [scalar] (phistar);
                    (Gamu) -- [gluon] (v1);
                    (Gbnu) -- [gluon] (v2);
                    (v1) -- [fermion, line width=\lwR] (Yuk2) -- [fermion, line width=\lwR] (vert) -- [fermion,line width=\lwR] (v2) -- [fermion,line width=\lwR] (Yuk1) -- [fermion,line width=\lwR] (v1); 
                    (Yuk1) -- [gluon] (Yuk2);
                };
            \end{feynman}    
        \end{tikzpicture}
    \caption{}\label{fig:discarded2LQCD}
    \end{subfigure}
    \hfill
    \begin{subfigure}[t]{0.3\textwidth}\centering
        \begin{tikzpicture}
            \begin{feynman}[scale=0.75, transform shape]
              \vertex (vert) at (0,0) [square dot, scale=\sizesqdot, black] {};
               \vertex (phi) at (1.5, .9375) {\(H\)};
               \vertex (phistar) at (1.5, -.9375) {\(H^*\)};
               \vertex (Gamu) at (-3.5, 1.25) {\(g \)};
               \vertex (Gbnu) at (-3.5, -1.25) {\( g \)};
               \vertex (v1) [dot, scale = \sizedot ] at (-2, 1.25) {};
               \vertex (v2) [dot, scale = \sizedot ] at (-2, -1.25) {};
               \vertex (Yuk1)[dot, scale = \sizedot ] at (-2, 0.) {};
               \vertex (Yuk2)[dot, scale = \sizedot ] at (-1., .625) {};
                \diagram*{
                    (phi) -- [scalar] (vert) -- [scalar] (phistar);
                    (Gamu) -- [gluon] (v1);
                    (Gbnu) -- [gluon] (v2);
                    (v1) -- [fermion, line width=\lwL] (Yuk2) -- [fermion, line width=\lwR] (vert) -- [fermion,line width=\lwR] (v2) -- [fermion,line width=\lwR] (Yuk1) -- [fermion,line width=\lwL] (v1); 
                    (Yuk1) -- [scalar] (Yuk2);
                };
            \end{feynman}    
        \end{tikzpicture}
      \caption{}\label{fig:discarded2LYukawa}
    \end{subfigure}
    \hfill
\caption{Sample of the diagrams with a single insertion of the operator $\op{Ht}{}$ which vanish because of kinematics, see main text for details. We use a thick (thin) line to denote a left-(right-)handed field, a dashed line to denote the Higgs doublet, a black square to denote an insertion of $\op{Ht}{}$ and a black dot to denote a SM vertex insertion. }  \label{fig:discardedDiagrams}
\end{figure}

\begin{figure}
\centering
    \begin{subfigure}[t]{0.3\textwidth}\centering
        \begin{tikzpicture}
            \begin{feynman}[scale=0.75, transform shape]
                \vertex (Gamu) at (-1.5,0) {\(g \)};
                \vertex (v1) [dot, scale = \sizedot] at (0,0) {};
                \vertex (v2) [dot, scale = \sizedot] at (2.5,0) [square dot, scale = \sizesqdot, black] {};
                \vertex (phi) at (4,0) {\(H\)};
                \vertex (v3) [dot, scale = \sizedot] at (2.5,-2.5) {};
                \vertex (phistar) at (4,-2.5) {\(H^*\)};
                \vertex (v4) [dot, scale = \sizedot] at (0,-2.5) {};
                \vertex (Gbnu) at (-1.5, -2.5) {\(g \)};
                \vertex (yuk) [dot, scale = \sizedot] at (0, -1.25) {};
                \diagram*{
                    (Gamu) -- [gluon] (v1);
                    (v2) -- [scalar] (phi);
                    (v1) -- [fermion] (v2) -- [fermion, line width=\lwR] (v3) -- [fermion, line width = \lwL] (v4) -- [fermion, line width = \lwL] (yuk) -- [fermion, line width=\lwR] (v1);
                    (phistar) -- [scalar] (v3);
                    (Gbnu) -- [gluon] (v4);
                    (yuk) -- [ scalar] (v2);
                };
            \end{feynman}    
        \end{tikzpicture}
    \caption{}\label{fig:Box2L}
    \end{subfigure}
    \hfill
 \begin{subfigure}[t]{0.3\textwidth}\centering
        \begin{tikzpicture}
            \begin{feynman}[scale=0.75, transform shape]
                \vertex (Gamu) at (-1.5,0) {\(g \)};
                \vertex (v1) [dot, scale = \sizedot] at (0,0) {};
                \vertex (v2) [dot, scale = \sizedot] at (2.5,0) [square dot, scale = \sizesqdot, black] {};
                \vertex (phi) at (4,0) {\(H\)};
                \vertex (v3) [dot, scale = \sizedot] at (2.5,-2.5) {};
                \vertex (phistar) at (4,-2.5) {\(H^*\)};
                \vertex (v4) [dot, scale = \sizedot] at (0,-2.5) {};
                \vertex (Gbnu) at (-1.5, -2.5) {\(g \)};
                \vertex (yuk) [dot, scale = \sizedot] at (1.25, 0) {};
                \diagram*{
                    (Gamu) -- [gluon] (v1);
                    (v2) -- [scalar] (phi);
                    (v1) -- [fermion, line width=\lwL](yuk)  -- [fermion, line width=\lwR] (v2) -- [fermion, line width=\lwR] (v3) -- [fermion, line width = \lwL] (v4) -- [fermion, line width = \lwL] (v1),
                    (phistar) -- [scalar] (v3),
                    (Gbnu) -- [gluon] (v4),
                    (yuk) -- [ scalar, half right] (v2);
                };
            \end{feynman}    
        \end{tikzpicture}
    \caption{}\label{fig:Box2LVertex}
    \end{subfigure}
    \hfill
    \begin{subfigure}[t]{0.3\textwidth}\centering
        \begin{tikzpicture}
            \begin{feynman}[scale=0.75, transform shape]
                \vertex (Gamu) at (-1.5,0) {\(g \)};
                \vertex (v1) [dot, scale = \sizedot] at (0,0) {};
                \vertex (v2) [dot, scale = \sizedot] at (2.5,0) [square dot, scale = 0.01, white] {};
                \vertex (phi) at (4,0) {\(H\)};
                \vertex (v3) [dot, scale = \sizedot] at (2.5,-2.5) {};
                \vertex (phistar) at (4,-2.5) {\(H^*\)};
                \vertex (v4) [dot, scale = \sizedot] at (0,-2.5) {};
                \vertex (Gbnu) at (-1.5, -2.5) {\(g \)};
                \diagram*{
                    (Gamu) -- [gluon] (v1);
                    (v2) -- [scalar] (phi);
                    (v1) -- [fermion, line width=\lwL] (v2) -- [fermion, line width=\lwR] (v3) -- [fermion, line width = \lwL] (v4) -- [fermion, line width = \lwL] (v1);
                    (phistar) -- [scalar] (v3);
                    (Gbnu) -- [gluon] (v4);
                };
            \end{feynman}  
            \node[shape=star,star points=5,star point ratio = 2,fill=white, draw,scale = 0.5,opacity=1] at (v2) {};
        \end{tikzpicture}
    \caption{}\label{fig:Box1L}
    \end{subfigure}
\caption{Sample of the two-loop order diagrams with a single insertion of the operator $\op{Ht}{}$.  We use a thick (thin) line to denote a left-(right-)handed field, a dashed line to denote the Higgs doublet, a black square to denote an insertion of $\op{Ht}{}$, a black dot to denote a SM vertex insertion and a white star to denote the Yukawa coupling counterterm.  }  \label{fig:BoxDiagrams}
\end{figure}
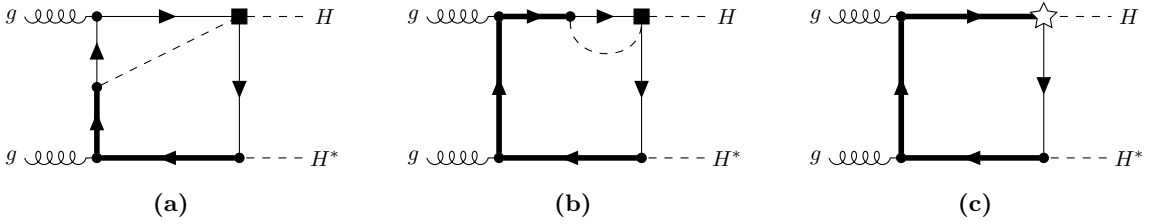

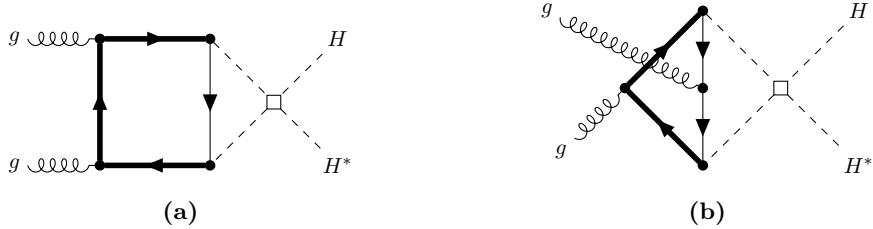
\begin{figure}
\centering
    \begin{subfigure}[t]{0.45\textwidth}\centering
        \begin{tikzpicture}
            \begin{feynman}[scale=0.75, transform shape]
                \vertex (g1) {\(g \)};
                \vertex (gtt1) [right=of g1, dot, scale=\sizedot] {}; 
                \vertex (htt1) [right= 55 pt of gtt1, dot, scale=\sizedot] {};
                \vertex (CH) [below right = 45 pt of htt1, square dot, scale=\sizesqdot, white] {};
                \vertex (htt2) [below left = 45 pt of CH, dot, scale = \sizedot] {};
                \vertex (gtt2) [left= 55 pt of htt2, dot,scale=\sizedot] {};
                \vertex (g2) [left of = gtt2] {\(g \)};
                \vertex (H1) [above right = 45 pt of CH] {$H$};
                \vertex (H2) [below right = 45 pt of  CH] {$H^*$};
                \diagram*{
                    (g1) -- [gluon] (gtt1);
                    (g2) -- [gluon] (gtt2);
                    (gtt1) -- [fermion, line width=\lwL] (htt1) -- [fermion, line width=\lwR] (htt2) -- [fermion, line width = \lwL] (gtt2) -- [fermion, line width = \lwL] (gtt1) ; 
                    (htt1) --[scalar] (CH) --[scalar] (htt2),
                    (H1) -- [scalar] (CH) -- [scalar] (H2);
                };
            \end{feynman}    
             \node[shape=regular polygon,regular polygon sides=4, aspect=1,fill=white, draw=black,scale = .5,opacity=1.] at (CH) {};  
        \end{tikzpicture}
    \caption{}\label{fig:Box2L_Class3}
    \end{subfigure}
  \begin{subfigure}[t]{0.45\textwidth}\centering
        \begin{tikzpicture}
            \begin{feynman}[scale=0.75, transform shape]
                \vertex (g1) {\(g \)};
                \vertex (gtt1) [above right=45 pt of g1, dot, scale=\sizedot] {}; 
                \vertex (htt1) [above right=55 pt of gtt1, dot, scale=\sizedot] {};
                \vertex (CH) [below right = 55 pt  of htt1, square dot, scale=\sizesqdot, white] {};
                \vertex (htt2) [below left = 55 pt  of CH, dot, scale = \sizedot] {};
                \vertex (gtt2) [right= 38.89 pt of gtt1, dot,scale=\sizedot] {};
                \vertex (g2) [above left= 55 pt of gtt1] {\(g \)};
                \vertex (H1) [above right = 55 pt of CH] {$H$};
                \vertex (H2) [below right = 55 pt of CH] {$H^*$};
                \diagram*{
                    (g1) -- [gluon] (gtt1);
                    (g2) -- [gluon] (gtt2);
                    (gtt1) -- [fermion, line width=\lwL] 
                    (htt1) -- [fermion, line width=\lwR]
                    (gtt2) -- [fermion, line width=\lwR](htt2) -- [fermion, line width = \lwL] (gtt1); 
                    (htt1) --[scalar] (CH) --[scalar] (htt2),
                    (H1) -- [scalar] (CH) -- [scalar] (H2);
                };
            \end{feynman}    
             \node[shape=regular polygon,regular polygon sides=4, aspect=1,fill=white, draw=black,scale = .5,opacity=1.] at (CH) {};  
        \end{tikzpicture}
    \caption{}\label{fig:Triangle2L_Class3}
    \end{subfigure}    
\caption{Sample of the two-loop order diagrams with a single insertion of the operators $\op{H\Box}{}$ and $ \op{HD}{}$.  We use a thick (thin) line to denote a left-(right-)handed field, a dashed line to denote the Higgs doublet, a white square to denote an insertion of $\op{H\Box}{}$ and $\op{HD}{}$ and a black dot to denote a SM vertex insertion.  }  \label{fig:DiagramsClass3}
\end{figure}

We describe in this section the computation of the two-loop contributions of the operators in Eqs.~\eqref{eq:psi2phi2Dops_phiqL}-\eqref{eq:psi2phi2Dops_phiud} and Eq.~\eqref{eq:phi4d2} to the RGE of $\coeff{HG}{}$. Practically, this requires the computation of the divergent parts of the two-loop corrections to a process where $\op{HG}{}$, defined in Eq.~\eqref{eq:OHG}, enters at tree-level. We perform our computation in the unbroken phase and we consider the process $gg \to H^* H$. 
In dimensional regularization, the divergences are parametrized by poles in $\epsilon=0$, being $D = 4 - 2\epsilon$ the number of space-time dimensions. 

We will discuss here in detail the calculation of the terms proportional to $\op{Ht}{}$. All the other operators of class 7 are obtained in analogy. 
At one-loop order, the contributions to $gg \to H^* H$ can only arise via triangle diagrams as in Fig.~\ref{fig:discarded1L}. However,  once all the relevant diagrams are considered their contribution vanishes \cite{rge3}. The reason is that the diagrams are purely $s$-channel, depending on the linear combination $q_1 + q_2=p_1+p_2$ where we have used $p_{1,2}$ ($q_{1,2}$) to denote the incoming (outgoing) momenta of the gluons (Higgs doublets). The scalar current gives rise to a contraction with $(p_1-p_2)$, yielding a vanishing result. 

Following this argument, we can safely ignore all the two-loop diagrams where the two external Higgs bosons are connected to the effective vertex. This applies both to pure QCD (see Fig.~\ref{fig:discarded2LQCD}) and to mixed Yukawa-QCD (see Fig.~\ref{fig:discarded2LYukawa}) contributions.\footnote{As a consequence of this statement, the class 7 pure QCD contribution to the RGE of $\coeff{HG}{}$ vanishes at all loop orders at dimension-six level. } 
We have cross-checked the correctness of our reasoning by explicitly computing the $\order{g_s^4}$ diagrams with a QCD correction to the propagator connecting the two QCD vertices and verifying that their sum vanishes after a trivial change of variable in the momentum integration. 

Our argument allows to restrict our attention to a total of 24 diagrams containing an insertion of $\op{Ht}{}$, where one of the two Higgs doublets in the EFT vertex is external and the other is internal. A sample of the two families of diagrams is displayed in Figs.~\ref{fig:Box2L} and \ref{fig:Box2LVertex}.
Moreover, class 7 operators correct the Yukawa coupling at one-loop \cite{rge1}, calling for the inclusion of the one-loop insertions of the one-loop Yukawa counterterms as in Fig.~\ref{fig:Box1L}.
Class 3 operators modify the quadrilinear coupling, yielding 12 diagrams, of which we present a sample in Fig.~\ref{fig:DiagramsClass3}.

We have performed two independent computations of the class 7 operator contributions to check the set-up of our computation.  
The first one relies on \texttt{FeynRules}~\cite{Alloul:2013bka, Christensen:2008py} for the generation of the Feynman rules, where the model file is obtained by extending the SM model file in the unbroken phase provided in the additional material to \texttt{MatchMakerEFT}~\cite{Carmona:2021xtq}. The resulting Feynman rules are presented in App.~\ref{app:feynmanrules}.
\texttt{FeynArts}~\cite{Hahn:2000kx} and \texttt{FeynCalc}~\cite{Shtabovenko:2023xyz,Shtabovenko:2020gxv, Shtabovenko:2016sxi,Mertig:1990an}, interfaced via \texttt{FeynHelpers} \cite{Shtabovenko:2016whf}, are used to generate the diagrams and to simplify the Dirac algebra. The numerical integration of the two-loop integrals is performed with \texttt{AMFlow}~\cite{Liu:2022chg} supplemented with \texttt{Blade}~\cite{Guan:2024byi}.

The second computation relies on the same pipeline of Ref.~\cite{DiNoi:2024ajj}. The diagrams are generated with {\tt qgraf-3.6.10} \cite{Nogueira:1991ex} and simplified with \texttt{FeynCalc}. The numeric integration is performed with \texttt{AMFlow} supplemented with \texttt{LiteRed2} \cite{Lee:2012cn,Lee:2013mka} and \texttt{FiniteFlow} \cite{Peraro:2019svx}.

We have checked that, when the divergences arising from the SMEFT one-loop contributions to the SM parameters \cite{rge1} are properly subtracted, the leftover divergence is a rational number, in units of $1/\epsilon \times 1/(16 \pi^2)^2$, up to the considered numeric precision of 16 digits. We have checked that our result is stable upon variation of the kinematic point.  

Both computations use the Na\"ive Dimensional Regularization (NDR) \cite{Bardeen:1972vi, CHANOWITZ1979225} scheme for the treatment of $\gamma_5$ in a non-integer number of dimensions. This prescription is known to be algebraically inconsistent in presence of traces involving an odd number of $\gamma_5$ in combination with at least six $\gamma$ matrices as in the diagrams we compute. However, we are interested in the contribution to the RGE of $\coeff{HG}{}$, meaning that we can ignore the contribution of the traces involving $\gamma_5$, which would renormalize its CP-odd counterpart, namely $\tilde{\mathcal{C}}_{HG}$.  

The Breitenlohner-Maison-t'Hooft-Veltman (BMHV) scheme \cite{Breitenlohner:1977hr, THOOFT1972189} 
has recently gained popularity as an alternative to the NDR scheme, being the only known continuation scheme proved to be algebraically consistent
\cite{Speer:1974cz,Breitenlohner:1975hg, Breitenlohner:1976te,Aoyama:1980yw,Costa:1977pd}. However, it highly increases the algebraic difficulty of loop computation. 
We have checked that, for the present computation, no difference arises between the two schemes by employing the map presented in Ref.~\cite{DiNoi:2025uan}. This procedure is legit in this context since the leading divergence is of $\order{1/\epsilon}$, as recently showcased in Ref.~\cite{DiNoi:2025arz}.

We note that, in principle, all the operators considered in this work can renormalize redundant operators, for which we follow the notation of Ref.~\cite{Carmona:2021xtq}. This renormalization arises from subamplitudes with schematic structure $\bar{\psi}\psi \to H^*H$, $\psi=Q_L,t_R,b_R$, which affect the operators $\mathcal{R}_{H\psi}',~\mathcal{R}_{H\psi}''$,\footnote{When $\psi=Q_L$, two possible isospin contractions are possible, generating a total of four redundant operators involving the left-handed quarks.} and from $H \to \bar{Q}_L \chi_R$, $\chi=b,t$ (possibly in association with one or two gluons), which impact the operators $\mathcal{R}_{uHDi}$, $i=1,2,3,4$. Since some of the external states in these subprocesses are not on-shell, the relevant one-loop insertions of redundant operators must be included in the total contribution to $gg \to H^*H$. The operators $\mathcal{R}_{H\psi}',~\mathcal{R}_{H\psi}''$ are potentially renormalized by the operators in class 3 at $\order{\yuk{}^2}$, but their one-loop contribution to the process $gg \to H^* H$ vanishes. Conversely, the one-loop insertion of $\mathcal{R}_{uHDi}$, $i=1,2,3,4$ does not vanish identically. These operators are renormalized by all the class 7 operators considered in our work, giving rise to potential non-vanishing contribution arising from the one-loop insertion of the associated one-loop counterterms. 
We have explicitly checked that, when the counterterms are expressed in terms of the WCs of the class 7 operators, these contributions to the full process $gg \to H^* H$ vanish. We have performed this check in the case where only $\op{Ht}{}$ is present, as representative for the class 7.  

Moreover, we do not need to consider contributions coming from one-particle reducible diagrams to $\op{HG}{}$. This conclusion stems from the fact that no redundant operator is reduced to $\op{HG}{}$ when the SM equations of motions are used \cite{dim6smeft}.

We obtain 

\begin{equation} \label{eq:CHGRGE}
\begin{aligned}
 \mu \frac{d \coeff{HG}{}}{d \mu}  \supset \frac{g_s^2}{(16 \pi^2)^2} \Big[ &
 +\yuk{b} \yuk{b}^* (\coeff{Hb}{}-\coeff{HQ}{(1)}-3 \coeff{HQ}{(3)})+\yuk{t} \yuk{t}^* (\coeff{HQ}{(1)}-3 \coeff{HQ}{(3)}-\coeff{Ht}{}) \\ &+ (\coeff{Htb}{} \yuk{b} \yuk{t}^*+\coeff{Htb}{*} \yuk{b}^* \yuk{t}) \\
 & -\left(\coeff{H\Box}{}-\frac{1}{2} \coeff{HD}{}\right)\left(\yuk{b} \yuk{b}^*+\yuk{t} \yuk{t}^* \right)  \qquad   \\ 
 & +\frac{3}{2} \left[ \coeff{t H}{} \yuk{t}+\coeff{t H}{*} \yuk{t}^* +\coeff{b H}{} \yuk{b}+\coeff{b H}{*} \yuk{b}^* \right]  \\ 
&- 4 \, \delta_{NDR} \left(   
\yuk{t}\yuk{t}^{*}   \left(\coeff{Qt}{(1)} - \frac{1}{6} \coeff{Qt}{(8)}  \right)
+{\yuk{b}\yuk{b}^{*}   \left(\coeff{Qb}{(1)} - \frac{1}{6} \coeff{Qb}{(8)}  \right)}
\right)
 \Big],
\end{aligned}
\end{equation}

\begin{equation}
\delta_{NDR} =
\begin{cases}
1 & \text{(NDR)}\\
0 & \text{(BMHV)}.
\end{cases}
\label{eq:deltaNDR}
\end{equation}
We use the same conventions as the one-loop results of \cite{rge1,rge2,rge3}.
The first two lines represent our novel result, while the contributions generated by four-quark operators and Yukawa-like operators (see Eq.~\eqref{eq:def_class5}) have been first derived in Ref.~\cite{DiNoi:2023ygk} and in Ref.~\cite{DiNoi:2024ajj},\footnote{In the previous expression, the contribution from the operators $\op{t H}{}$ and $\op{b H}{}$ presents an extra 1/2 with respect to the result in previous versions of Ref.~\cite{DiNoi:2024ajj}. The discrepancy has been fixed in the most recent version of reference. } respectively, which we point to for the precise definition of the operators. We stress that only the four-top contribution depends on the $\gamma_5$ scheme, while the others do not. 
\par
The final amplitude cannot exhibit infrared (IR) divergences, since no corresponding real-emission diagrams exist at the considered order in the coupling constants. However, IR divergences can still appear at intermediate stages of the calculation. As they must ultimately cancel, one has to verify that this cancellation does not occur against ultraviolet (UV) divergences, as can happen, for instance, in the case of scaleless integrals.

By replacing the gluons with electroweak gauge bosons in the diagrams in Figs.~\ref{fig:BoxDiagrams},\ref{fig:DiagramsClass3} we could obtain a similar result for the operators  $\op{HW}{},\, \op{HB}{},\,\op{HWB}{}$ (defined in Eqs.~\eqref{eq:def_class4_HWHB}--\eqref{eq:def_class4_HWB}), as in Ref.~\cite{Haisch:2025vqj}. 
However, in our case it would represent an incomplete result, since in this scenario additional diagrams arise as a consequence of the non-vanishing $\mathrm{SU(2)_L \times U(1)_{Y}}$ charge of the Higgs doublet.

Obtaining the RGEs for such operators is hence beyond the scope of this paper.
The phenomenological relevance of these operators lies in their impact in the interactions involving photons or a photon and $Z$ bosons and (at least) one Higgs boson and they lead to a shift in the electroweak input \cite{Biekotter:2023xle}. 
In the SM, such interactions are numerically dominated by $W$ loops, so we expect the two-loop running contributions to be less numerically important than in the gluon case such that the phenomenological study we perform in the next section remains valid omitting these sub-leading contributions.  Furthermore, their contributions to the Higgs couplings to massive vector bosons is suppressed in the loop counting with respect to the tree-level contributions from class 3 operators.

\section{Phenomenology} \label{sec:pheno}

In this section, we study the effect of the inclusion of two-loop running effects with fits to Higgs data at the LHC. In particular, we focus on the modifications to the signal strengths of a set of Higgs decay final states due to the presence of SMEFT operators.

We perform two different fits. In first instance, we adopt a bottom-up perspective by performing the analysis in terms of the WCs to observe how their bounds are modified by the inclusion of the running. Secondly, we consider a top-down approach by selecting UV models with pairs of Vector-like Quarks (VLQs) coupled to the third-generation SM quarks, which generate, at tree-level, at least one of the class 7 operators considered in this paper.

The WCs that enter the observables in this section are computed at the low scale $\mu_{R} = m_h = 125\, \mathrm{GeV}$.  Their expression in terms of the WCs at the high energy scale $\Lambda = 2 \, \mathrm{TeV}$ can be obtained with the evolution matrix formalism \cite{Fuentes-Martin:2020zaz} 
\begin{equation}\label{eq:evmat}
\coeff{i}{}(\mu_R) = U_{ij}(\mu_R,\Lambda) \coeff{j}{}(\Lambda).
\end{equation}
In the present case, the evolution matrix $U_{ij}(\mu_R,\Lambda)$ has been obtained with a modified version of \texttt{RGESolver} \cite{DiNoi:2022ejg} relying on a numeric solution of the RGEs. All the operators included in this section have been defined in Sec.~\ref{sec:notation}.

\paragraph{Cross-section} The SM value for the Higgs boson production cross section in the dominant gluon-gluon fusion contribution is $\sigma^{\rm SM} = \SI{48.68}{\pico\barn} $ at a center of mass energy of $\sqrt{s}=13\text{ TeV}$. In the SMEFT, the cross-section reads
\begin{equation}
  \sigma=  \left[48.68 + 2.83\cdot 10^4 v^2 C_{H G} +97.36 v^2 C_{H,\text{kin}}-112.25 \frac{v^2}{\yuk{t}} C_{tH}\right]\text{pb}\,,
\label{eq:HiggsProductionCrossSection}\end{equation}
where $\coeff{H,\text{kin}}{}$ was defined in Eq.~\eqref{eq:Hkin_definition}. We have obtained this parameteric dependence on the SMEFT WCs using the EFT implementation \cite{Grazzini:2018eyk} in \texttt{higlu} \cite{Spira:1995mt}.

\paragraph{Higgs total and partial decay widths} The Higgs total decay width in the SMEFT at LO in the U(3)$^5$ flavour-symmetric limit was computed in Ref.~\cite{Brivio:2019myy}. We adapt those results, noting that quark operators of class 7 contribute only via corrections to the gauge boson total width except for the cases with quarks in final states from $Z$ boson decays. In this case there are some direct contributions where the Higgs boson and one $Z$ boson couples via a class 7 operator to bottom quarks, and the $Z$ boson decays subsequently. This results in  
\begin{align}
    \Gamma_h = \Gamma_h^{\rm SM} & \left( 1 + 50.6\, v^2 \coeff{H G}{} - 1.50 \, v^2 \coeff{HB}{} -1.21 \,v^2 \coeff{HW}{} + 1.21 \, v^2 \coeff{HWB}{}+1.83 \, v^2 \coeff{H\Box}{} \nonumber \right.\\
    &  - 0.43\, v^2 \coeff{H D}{} -48.5\,v^2 \yuk{b}\,\coeff{bH}{}+0.00016 v^2 \,\coeff{Hq}{(1)}+0.0015 v^2\,\coeff{Hq}{(3)} -0.00023 v^2 \, \coeff{Hb}{}    \nonumber \\
    &\left.+ 2\,v^2 \left(\text{BR}_{h\rightarrow \gamma\gamma} \left[\coeff{H, \text{kin} }{} + \frac{0.53}{ \yuk{t}} \coeff{tH}{} \right] + \text{BR}_{h\rightarrow gg} \left[ \coeff{H, \text{kin} }{} -\frac{2.12}{\yuk{t}} \coeff{tH}{}  \right]\right) \right)\,,
\label{eq:HiggsDecayWidth}\end{align}
using the $\{M_W, M_Z, G_F\}$ input scheme for the electroweak parameters.
The last line encodes the higher-order rescaling of the Higgs decays into photon and gluon pairs through $\coeff{H,\text{kin}}{}$ defined in Eq.~\eqref{eq:Hkin_definition}, together with the additional modifications induced by shifts in the top Yukawa coupling. In the UV models we consider, also the bottom Yukawa coupling can be modified but we do not consider its effect in the $h\to gg$ and $h\to \gamma\gamma$ decays as it is negligible with respect to the top Yukawa contribution. 
We note already from this formula that generically our two-loop contributions can be of the same order than the direct contributions. Indeed, since the $\coeff{HG}{}$ has a large pre-factor, even the two-loop factor $1/(16\pi^2)^2 \log(\Lambda/m_h)$ can be compensated.


We obtain the partial decay widths in the SMEFT rescaling the results in Ref.~\cite{Brivio:2019myy} to take into account the modifications of the decay into photon pairs and the top Yukawa coupling, analogously to the case of the total width discussed previously. 
We have validated these expressions to those reported in Ref.~\cite{Contino:2014aaa}.\footnote{Ref.~\cite{Contino:2014aaa} uses the SILH basis \cite{Giudice:2007fh}, whereas Ref.~\cite{Brivio:2019myy} uses the Warsaw basis. To compare the results, we translate both sets of results into the Greens' basis by means of Ref.~\cite{Gherardi:2020det} and compare the terms of interest.}

For the decays into fermion pairs, introducing $l\in \{\mu^-, \, \tau^-\}$ , the partial decay widths are 
\begin{align}
    \frac{\Gamma_{h\rightarrow l \bar{l}}}{\Gamma_{h\rightarrow l \bar{l}}^{\rm SM}} &=  1 + 2 v^2   \, \coeff{H,\text{kin}}
  \,, \label{eq:hll}\\
    \frac{\Gamma_{h\rightarrow b \bar{b}}}{\Gamma_{h\rightarrow b \bar{b}}^{\rm SM}} &=  1 + v^2 \left( 2 \coeff{H,\text{kin}}{}
    - 2\,\coeff{bH}{} \right)\,. \label{eq:partialwidth_bb}
\end{align}
For the decays into photon pairs we get 
\begin{equation}
    \frac{\Gamma_{h\rightarrow \gamma\gamma}}{\Gamma_{h\rightarrow \gamma\gamma}^{\rm SM}} = 1 + v^2 \left( 2 \, \coeff{H,\text{kin}}{}
    +\frac{0.53}{\yuk{t}} \, \coeff{tH}{} - 231 \,\coeff{HW}{}- 805\, \coeff{HB}{} +431\, \coeff{HWB}{} \right)\,.\label{eq:partialdecaywidth_gammagamma}
\end{equation}
\indent As for decays into pairs of $W$ and $Z$ pairs, we further consider their decay products in order to compare with experimental results. In particular, Ref.~\cite{CMS:2022dwd}, whose experimental results we will be using later, considers the processes $h \rightarrow W^+W^- \rightarrow 2l \, 2\nu$ and $h \rightarrow ZZ \rightarrow 4l$. In order to get the partial decay width for these processes, we start again from Ref.~\cite{Brivio:2019myy} and focus on their results for partial decay widths of the Higgs boson into four-fermion final states mediated by "only neutral", "only charged" and "neutral plus charged" currents. In our scenario, the first case reduces to the $4l$ final state, while the second and third correspond to the $2 l \, 2\nu$ final state. The SM partial decay width for $h\rightarrow ZZ \rightarrow 4l$ is then $\Gamma^{\rm SM}_{4l} = \SI{1.1}{\kilo\electronvolt}$, the partial decay width for $h \rightarrow W^+W^- \rightarrow 2l \, 2\nu$ reads $\Gamma_{2l2\nu}^{\rm SM} = \SI{90}{\kilo\electronvolt}$ , and the corresponding SMEFT expressions read 
\begin{align} \nonumber
    \frac{\Gamma_{4l}}{\Gamma_{4l}^{\rm SM}} =&1 + v^2 \left(2 \coeff{H\Box}{} + 0.27\,\coeff{HD}{} +0.17\,\coeff{HW}{}-1.66  \,\coeff{HB}{}-0.51\,\coeff{HWB}{} \right. \\ & \left. -0.27\,\coeff{HQ}{(1)} - 0.27 \, \coeff{HQ}{(3)}  + 0.048\,\coeff{Hb}{} \right)\,,  \label{eq:hllll}\\ \nonumber
   \frac{\Gamma_{2l2\nu}}{\Gamma_{2l2\nu}^{\rm SM}} =&1 + v^2 \left(2 \coeff{H\Box}{} -0.52 \,\coeff{HD}{} -1.5 \,\coeff{HW}{}+0.003 \, \coeff{HB}{}-0.023\,\coeff{HWB}{}\right. \\ & \left. -0.0046\,\coeff{HQ}{(1)} - 0.0046\, \coeff{HQ}{(3)}  + 0.00081\,\coeff{Hb}{} \right) \,.  \label{eq:hllnunu} 
\end{align}


\subsection{Fitting framework} \label{subsec:pheno_obsandfit}
Adopting the symbol $\mathcal{O}_\alpha$ to identify an observable $\alpha$, the $\chi^2$-function can be defined as
\begin{equation}
    \chi^2 = \sum_{\alpha,\,\beta} \left(\mathcal{O}_\alpha^{\rm exp} - \mathcal{O}_\alpha^{\rm th} \right) \left[C^{-2}\right]_{\alpha \beta} \left(\mathcal{O}_\beta^{\rm exp} - \mathcal{O}_\beta^{\rm th} \right)\,,
\end{equation}
where $C^{-2}$ is the inverse of the covariance matrix and the superscript 'th' ('exp') means that the theoretical (experimental) value of the observable $\mathcal{O}_\alpha$ is being considered. The theoretical value accounts for both the SM prediction and the deviations due to the insertion of SMEFT operators.

Following Ref.~\cite{Erdelyi:2024sls}, we construct a fit to Higgs data using the measurements of signal strengths with their covariance matrix as provided in Ref.~\cite{CMS:2022dwd}. Considering the Higgs decay channels $h \to \alpha$ with $\alpha\in \{W^+W^-, \, ZZ, \, b\bar{b}, \, \tau^+ \tau^-, \, \mu^+\mu^-, \, \gamma\gamma\}$,\footnote{For the decays into pairs of $W$ and $Z$ bosons, Ref.~\cite{CMS:2022dwd} considers the bosonic decay channels $ZZ\rightarrow 4l$ and $WW\rightarrow l\nu l\nu$. Thus, for the first two decay channels, we can use the decay widths of Eqs.~\eqref{eq:hllll} and \eqref{eq:hllnunu} respectively.} for each channel it is given by
\begin{equation}
    \mu_\alpha = \frac{\sigma \times \text{BR}_{h\rightarrow \alpha}}{\sigma^{\rm SM} \times \text{BR}^{\rm SM}_{h\rightarrow \alpha}} = \frac{\sigma}{\sigma^{\rm SM}} \frac{\Gamma_h^{\rm SM}}{\Gamma_{h\rightarrow \alpha}^{\rm SM}} \frac{\Gamma_{h\rightarrow \alpha}}{\Gamma_h} = \frac{\sigma}{\sigma^{\rm SM}} \frac{\Gamma_h^{\rm SM}}{ \Gamma_h } \frac{\Gamma_{h\rightarrow \alpha}}{ \Gamma_{h\rightarrow \alpha}^{\rm SM}}\,,
\label{eq:signalstrength_definition}\end{equation}
in which $\sigma^{(\rm SM)}$ and BR$^{( \rm SM)}$ denote, respectively, the value in the SMEFT (SM) of the Higgs production cross-section and the branching ratio into the specific channel. 
We consider up to order $\mathcal{O}(1/\Lambda^2)$ in the signal strength.
Of the three multiplied ratios in the final expression, the first is known from Eq.~\eqref{eq:HiggsProductionCrossSection}, the second from Eq.~\eqref{eq:HiggsDecayWidth} and the third is selected from Eqs.~\eqref{eq:hll}-\eqref{eq:hllnunu} depending on the decay channel of interest.

The allowed region satisifies the condition
\begin{equation}
    \chi^2 < \chi^2_{\rm min} + \Delta\chi^2\left(n , \text{CL}\right) \,.  
\label{eq:chi2_fit}\end{equation}
In the previous expression, $n$ denotes the degrees of freedom (number of couplings in the top-down approach, number of involved WCs in the bottom-up approach) while CL stands for Confidence Level.  The value $\chi^2_{\rm min}$  is the minimum of the $\chi^2$.

\subsection{Results}\label{subsec:pheno_fit}

We now present our results. We notice that the operators $\op{Htb}{}$ and $\op{Ht}{}$ do not enter directly any of the Eqs.~\eqref{eq:HiggsProductionCrossSection}-\eqref{eq:hllnunu}, so they enter only via mixing effects with other operators.

\subsubsection{Bottom-up results}\label{subsubsec:pheno_bottomup}
In this section we report our one- and two-parameter results for the bottom-up fits.
We stress that these parameters are set to be non-zero at the high-scale $\Lambda =\SI{2}{\tera\electronvolt}$, meaning that several other operators are generated via operator mixing, governed by Eq.~\eqref{eq:evmat}. Hence, the bounds are reported for the operators at such scale. 

\paragraph{One parameter fits}
\begin{figure}
    \centering
    \includegraphics[width=0.75\linewidth]{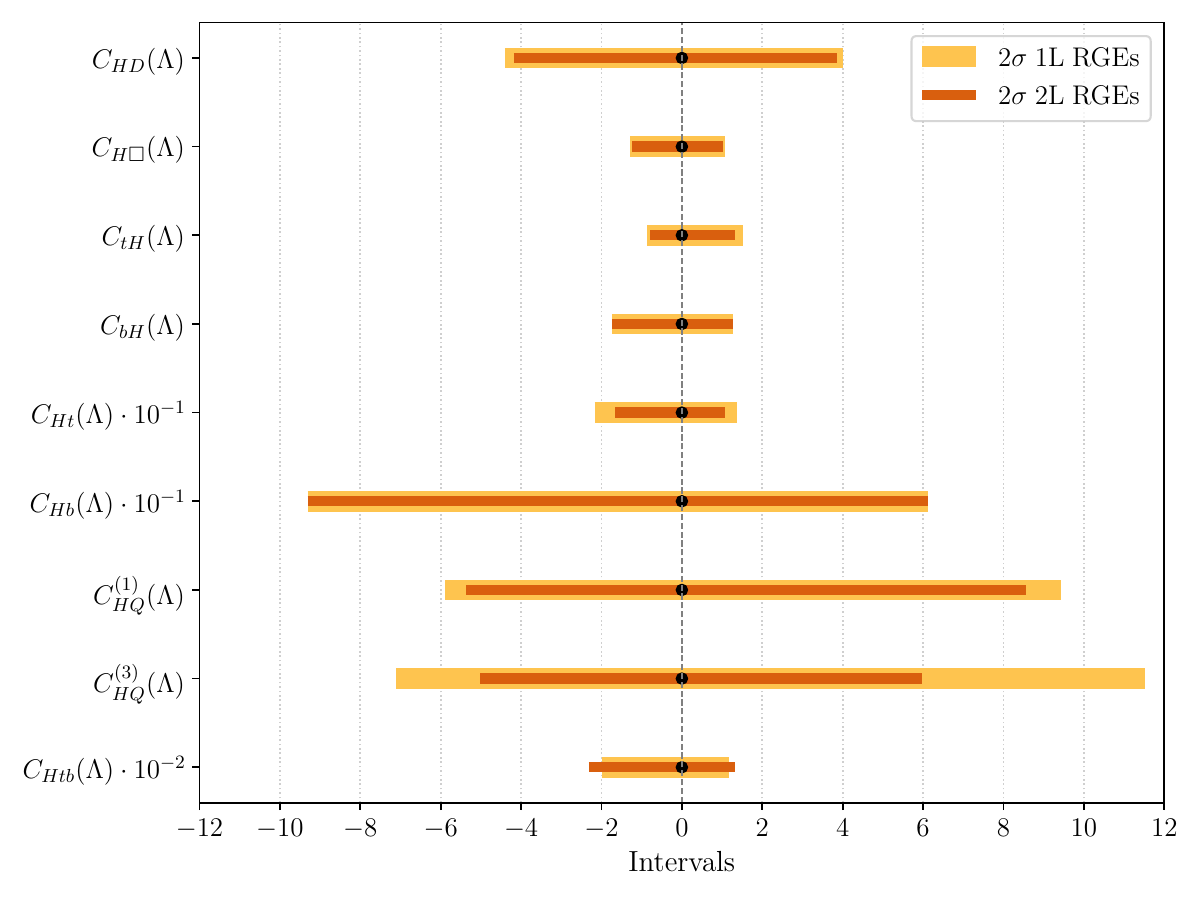}
    \caption{Allowed ranges at $2\sigma$ for the WCs obtained by performing one parameter fits. The yellow (orange) line is obtained by considering one-loop (two-loop) running effects. }\label{fig:results_oneparameter_1Lvs2L}
\end{figure}
We present the individual $2 \sigma$ CL interval for each WC considered at the energy scale $\Lambda=\SI{2}{\tera\electronvolt}$ in Fig.~\ref{fig:results_oneparameter_1Lvs2L}.
We note that the intervals for the class 3 coefficients are not sizeably affected by the inclusion of the two-loop effects in the RGEs. 
These operators are already constrained by direct contributions to the couplings with the Higgs bosons, reducing the impact of higher order corrections. 
Similar conclusions apply to the WCs of the Yukawa-like operators $\coeff{tH}{}$ and $\coeff{bH}{}$, which enter directly in the Higgs production cross section and in several decay channels. 

As for the class 7 operators, $\coeff{Hb}{}$ is not sizeably modified by the inclusion of the two-loop running effects, due to the bottom Yukawa suppression. Conversely, $\coeff{HQ}{(1)}$ and $\coeff{Ht}{}$ show a moderate reduction when two-loop running effects are accounted, while for $\coeff{HQ}{(3)}$ the reduction is larger. Moreover, we note that the bounds for $\coeff{Htb}{}$ are looser at two-loop level, hinting for some cancellations happening with respect to one-loop running effects.

Finally, we note that the bounds we present here are much weaker than the ones presented in Ref.~\cite{terHoeve:2025gey}. Since our focus is on the impact of the two-loop running effects in the fit, we only accounted for inclusive Higgs data and not, for instance, for electroweak precision data or top data as Ref.~\cite{terHoeve:2025gey}.\\
We note though that our results are still interesting in the light of constraints arising from the electroweak precision tests which are sensitive to class 7 operators and $\coeff{HD}{}$. For instance, the operator combination
\begin{equation}
    \coeff{HQ}{(-)} =    \coeff{HQ}{(1)} - \coeff{HQ}{(3)}\,. \label{eq:chqminus_def}
\end{equation}
exhibits a flat direction in the $Z$ pole measurements. This direction can be lifted using also $W$ boson data from the LEP $\SI{240}{\giga\electronvolt}$  \cite{Brivio:2017bnu, Celada:2024mcf} and di-boson data from the LHC. Nevertheless, it remains much less constrained, resulting in bounds of $\mathcal{O}(1/\Lambda^2)$ of $[-6.4,30.3]/\text{TeV}^2$ at a scale $\Lambda=\SI{1}{\tera\electronvolt} $ marginalised over the other parameters according to Ref.~\cite{terHoeve:2025gey}. As a comparison, for a single parameter fit on $\coeff{HQ}{(-)}$ we find
\begin{align}
    &\left[ -5.9 , 9.4  \right]/\text{TeV}^2 \, \quad\enspace \text{including 1L RGEs}\,,\\
    & \left[-5.4, 8.6 \right]/\text{TeV}^2  \,\quad \enspace \text{including 2L RGEs}\,.
\end{align}
To ease the comparison with the results of Ref.~\cite{terHoeve:2025gey}, in which the high-scale is set to be $\Lambda = \SI{1}{\tera\electronvolt}$, we recomputed the evolution matrix $U$ defined in Eq.~\eqref{eq:evmat} setting the  heavy scale to be at $\Lambda = \SI{1}{\tera\electronvolt}$. In this case our results from a single parameter fit on $\coeff{HQ}{(-)}$  are 
\begin{align}
    &\left[ -6.4 , 10.2  \right]/\text{TeV}^2 \, \quad\enspace \text{including 1L RGEs}\,,\\
    & \left[-5.9, 9.4 \right]/\text{TeV}^2  \,\quad \enspace \text{including 2L RGEs}\,.
\end{align}

\paragraph{Two parameter fits}
\begin{figure}[t]
    \centering
    \begin{subfigure}[b]{0.46\textwidth}
        \centering
        \includegraphics[width=\textwidth]{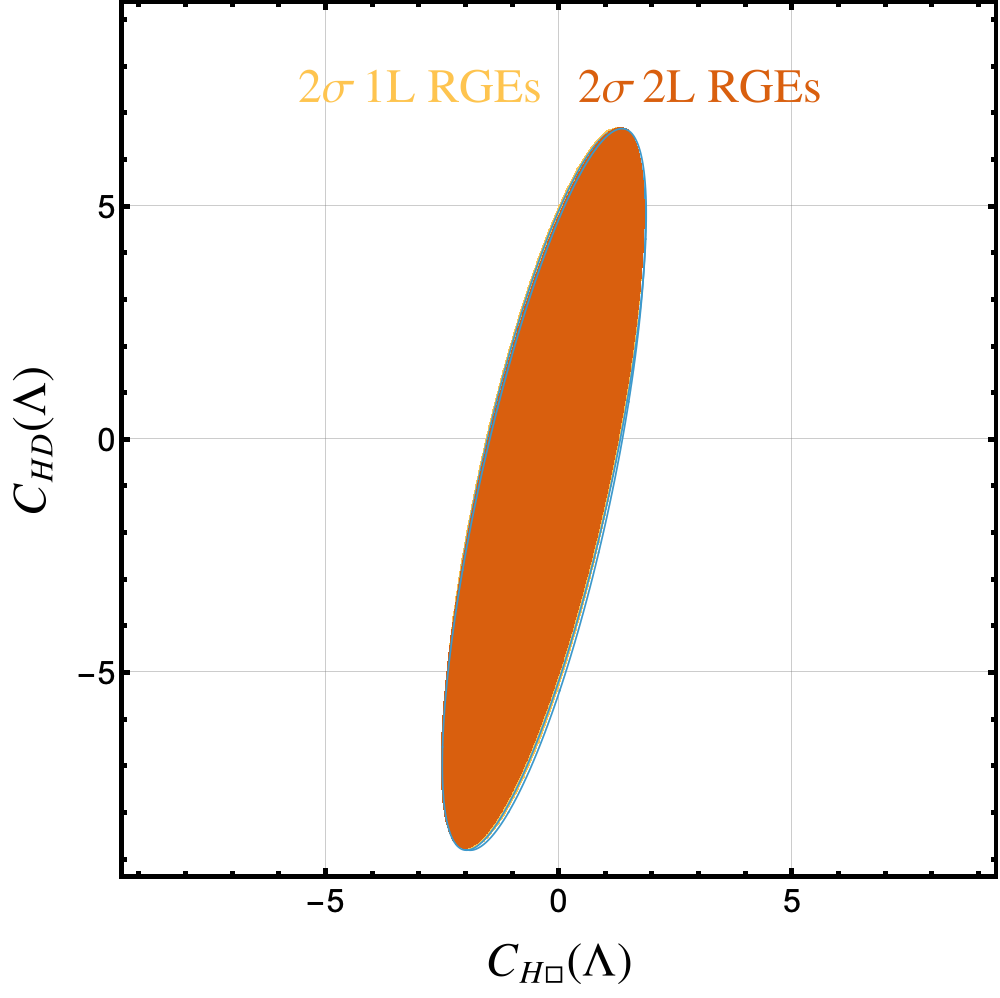}
        \caption{Class 3 operators} \label{fig:pheno_bottomup_twoparameterfits_HDHbox}
    \end{subfigure}
    \hfill 
    \begin{subfigure}[b]{0.47 \textwidth}
        \centering
        \includegraphics[width = \textwidth]{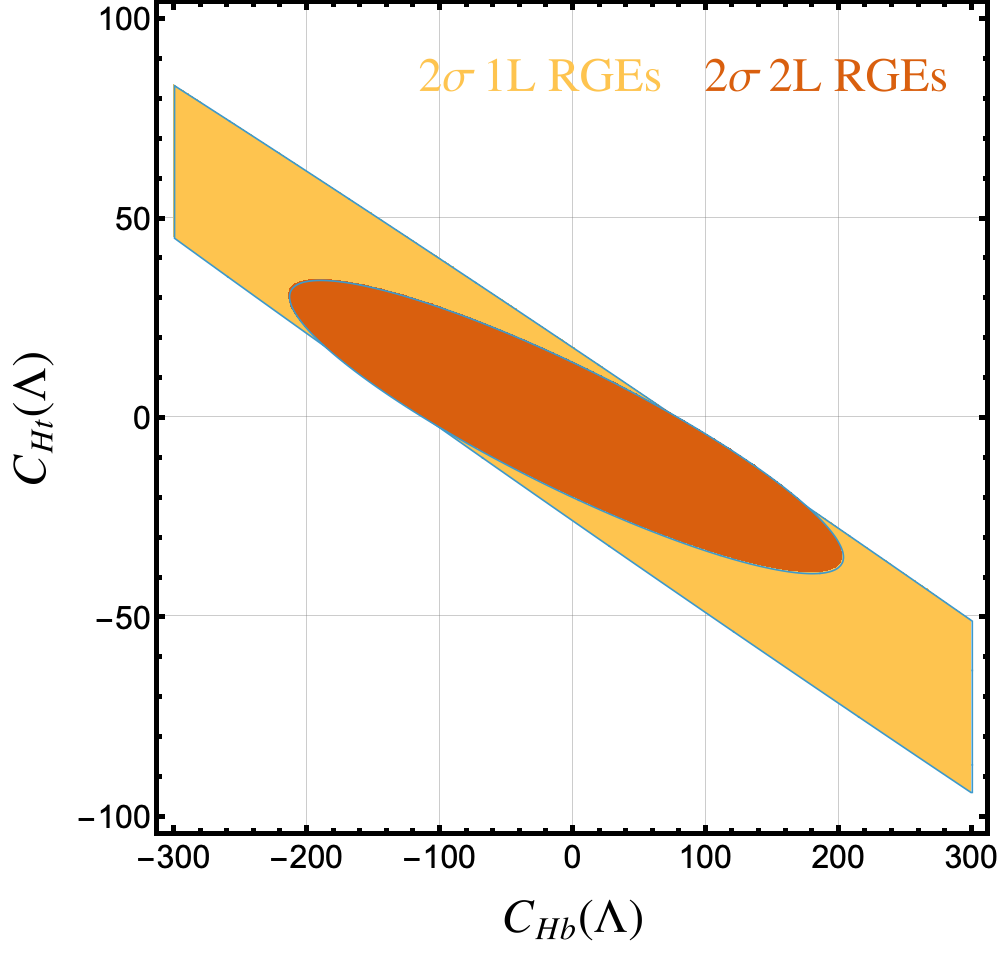}
        \caption{Class 7 operators with quark singlets}\label{fig:pheno_bottomup_twoparameterfits_HuHd}
    \end{subfigure}        
        \\
    \vspace*{1.1em}
     \begin{subfigure}[b]{0.46\textwidth}
        \centering
        \includegraphics[width = 1.\textwidth]{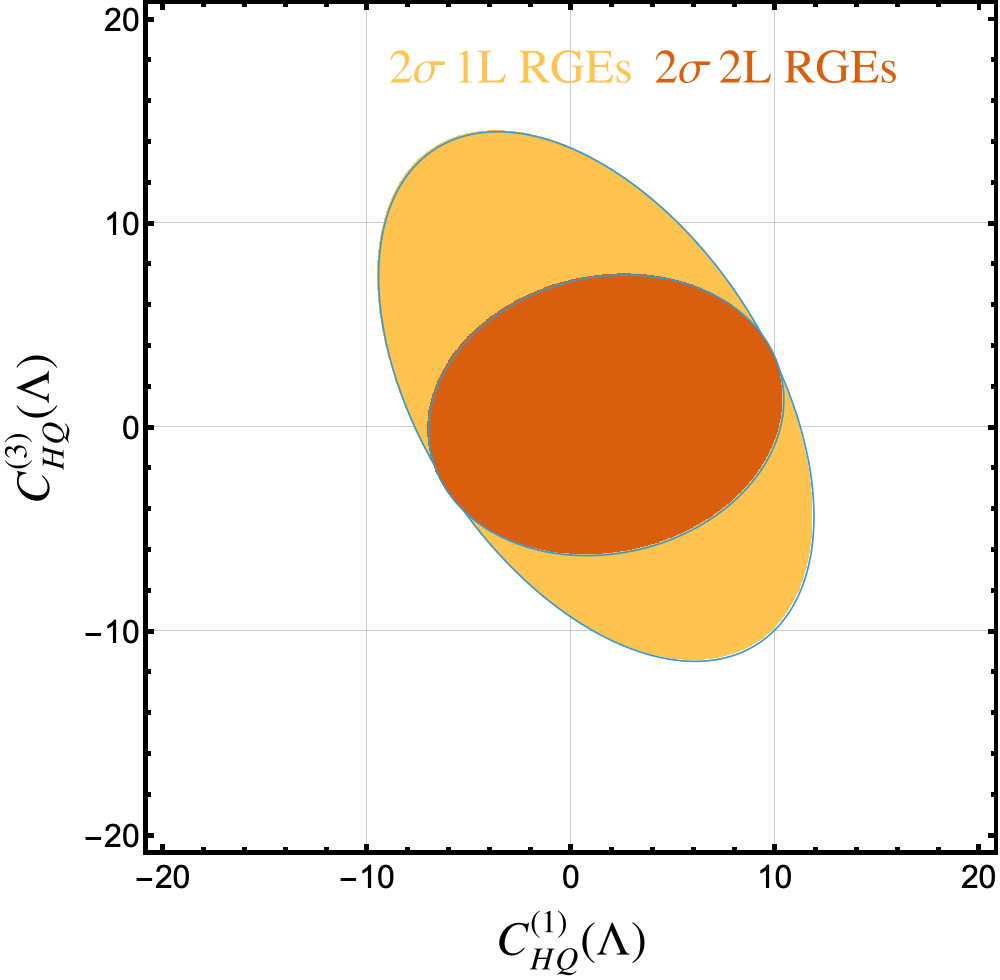}
        \caption{Class 7 operators $\coeff{HQ}{(1)}$ and $\coeff{HQ}{(3)}$}\label{fig:pheno_bottomup_twoparameterfits_Hq1Hq3}
    \end{subfigure}
    \hfill
    \begin{subfigure}[b]{0.46\textwidth}
        \centering
        \includegraphics[width = \textwidth]{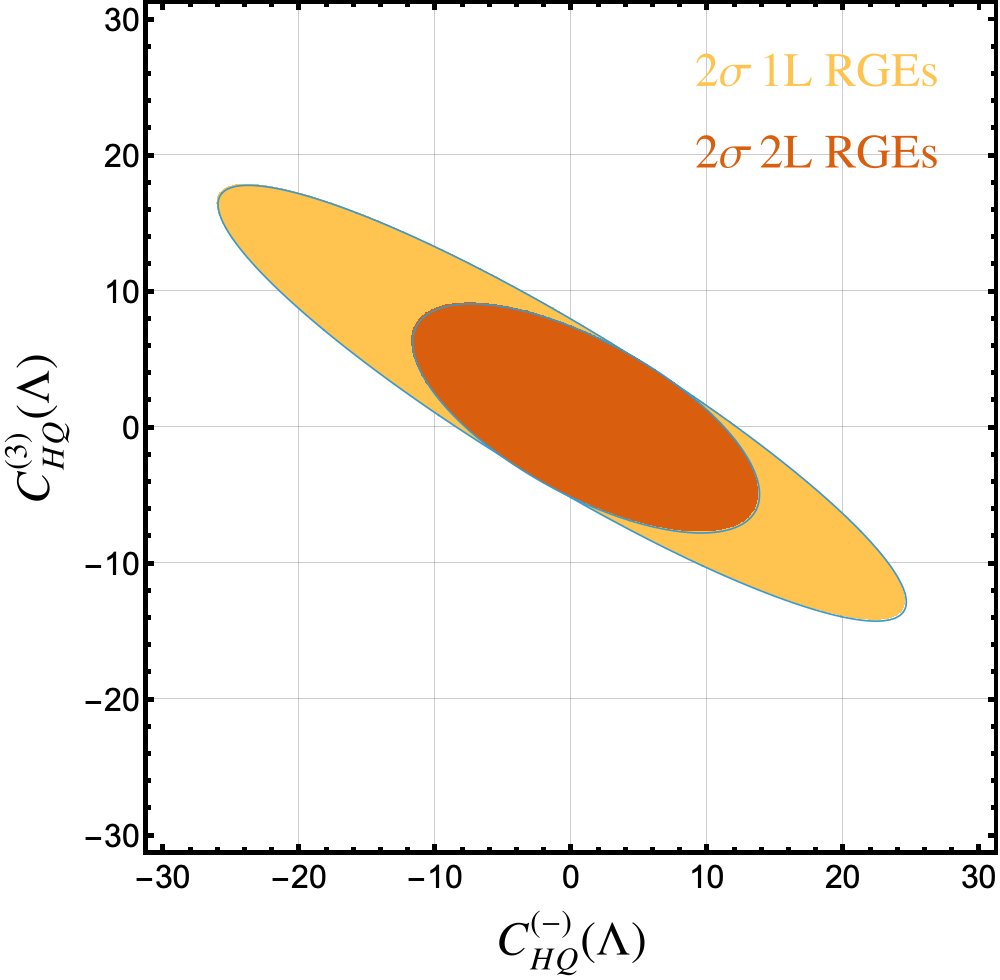}
        \caption{Class 7 operators $\coeff{HQ}{(-)}$ and $\coeff{HQ}{(3)}$ , $\Lambda=\SI{1}{\tera\electronvolt}$}\label{fig:pheno_bottomup_twoparameterfits_HqminusvsHq3}
    \end{subfigure}
    \caption{Allowed parameter spaces at $2\sigma$, obtained through a two parameter fit including one- (two-) loop running effects, are shown in yellow (orange). We consider in pairs the operators belonging to class 3 and to class 7.}
    \label{fig:pheno_bottomup_twoparameterfits}
\end{figure}
In Fig.~\ref{fig:pheno_bottomup_twoparameterfits} we allow two WCs to be non-vanishing at the energy scale $\Lambda = \SI{2}{\tera\electronvolt}$.   We present the results of four two-parameter fits, comparing the impact of including only one-loop or two-loop running effects. In Fig.~\ref{fig:pheno_bottomup_twoparameterfits_HDHbox}, the two class 3 operators are considered. The correlation between the two arises from the fact that their contribution appears mostly in the linear combination $\coeff{H,\, \text{kin}}{} = \coeff{H\Box}{} - \frac{1}{4}\coeff{HD}{}$.
Moving on to Fig.~\ref{fig:pheno_bottomup_twoparameterfits_HuHd} we compare $\coeff{Ht}{}$ and $\coeff{Hb}{}$. In this case, there is a significant improvement in the bounds considering that  $\coeff{Ht}{}$ does not enter any of the observables defined in Eqs.~\eqref{eq:HiggsProductionCrossSection}--\eqref{eq:hllnunu}, while $\coeff{Hb}{}$ enters but is very suppressed compared to the other WCs.\\
Finally, we present two plots (Fig.~\ref{fig:pheno_bottomup_twoparameterfits_Hq1Hq3} and Fig.~\ref{fig:pheno_bottomup_twoparameterfits_HqminusvsHq3}) involving class 7 operators with quark doublets, which, in motivated UV extensions of the SM, appear together, as shown in Tab.~\ref{tab:VLQs_treelevelops}. While in Fig.~\ref{fig:pheno_bottomup_twoparameterfits_Hq1Hq3} we study the bounds on the Warsaw basis pair $\{\coeff{HQ}{(1)}, \coeff{HQ}{(3)}\}$, in Fig.~\ref{fig:pheno_bottomup_twoparameterfits_HqminusvsHq3} we perform a coefficient redefinition and consider the allowed parameter space for the pair $\{\coeff{HQ}{(-)}, \coeff{HQ}{(3)} \}$ (with $\coeff{HQ}{(-)}$ defined in Eq.~\eqref{eq:chqminus_def}) and heavy scale set to be $\Lambda = \SI{1}{\tera\electronvolt}$. This redefinition eases the comparison with the results by global fits such as the ones in Ref.~\cite{terHoeve:2025gey}. Independently from the choice of parameters, in both cases the inclusion of two-loop running effects improves the constraints.

\subsubsection{Top-down results}\label{subsubsec:pheno_topdown}
\begin{table}[t]
    \centering
    {\renewcommand{\arraystretch}{1.7}
    \begin{tabular}{|c|c|c|c|c|c|c|}
    \hline
        & $\mathcal{G}_{\rm SM}$ & $\op{HQ}{(1)}$ & $\op{HQ}{(3)}$ & $\op{Ht}{}$ & $\op{Hb}{}$ & $\op{Htb}{}$ \\
    \hline
    $U$ & $\left(3,1\right)_\frac{2}{3}$ & \checkmark & \checkmark & & & \\
    \hline
    $Q_1$ & $\left(3,2\right)_\frac{1}{6}$ & & & \checkmark & \checkmark & \checkmark \\
    \hline
    $Q_7$ & $\left(3,2\right)_\frac{7}{6}$ & & & \checkmark & & \\
    \hline
    $T_1$ & $\left(3,3\right)_{-\frac{1}{3}}$ & \checkmark & \checkmark & & & \\
    \hline
    $T_2$ & $\left(3,3\right)_\frac{2}{3}$ & \checkmark & \checkmark & & & \\
    \hline
    \end{tabular}
    }
    \hfill{\renewcommand{\arraystretch}{1.7}
    \begin{tabular}{|c|c|}
    \hline
    Model & Particle Content \\
    \hline
    1     & $U + Q_1$ \\
    \hline 
    2     & $U + Q_7$ \\
    \hline 
    3     & $Q_1 + T_1$ \\
    \hline
    4     & $Q_1 + T_2$ \\
    \hline 
    5     & $Q_7 + T_2$ \\
    \hline
    \end{tabular}
    }
    \caption{Class 7 operators generated at the tree-level by the VLQs considered in this work, along with the charges under the SM gauge group of the VLQs, with $\mathcal{G}_{\rm SM} = \left(\text{SU}(3)_\text{C}, \, \text{SU}(2)_\text{L}\right)_{\text{U}(1)_\text{Y}}$ (left). The presented VLQs will be used to construct models with pairs of these particles (right). }
    \label{tab:VLQs_treelevelops}
\end{table}

In this second approach we consider UV models featuring Vector-like Quarks (VLQs) as they generate at tree level both the operators in Eqs.~\eqref{eq:psi2phi2Dops_phiqL}-\eqref{eq:psi2phi2Dops_phiud} and the Yukawa-like operators $\op{tH}{}$ and $\op{bH}{}$ \cite{deBlas:2017xtg}. 
The operators $\op{HG}{}$, $\op{HD}{}$ and $\op{H\Box}{}$ are instead generated at one-loop level.

In a similar spirit to Ref.~\cite{Erdelyi:2024sls}, we consider models with pairs of VLQs which give rise to unsuppressed contributions to the operator $\op{tH}{}$. 
In Tab.~\ref{tab:VLQs_treelevelops} we list the VLQs considered, their charges under the SM gauge group $\mathcal{G}_{\rm SM}$, the class 7 operators they generate at tree-level and how the five models considered in this work are defined. 

Given that we are interested in the RGE effects due to terms proportional to third generation Yukawa couplings, we make the simplifying assumption that the VLQs only couple to the third generation SM quarks. Furthermore, we set their masses to be $\Lambda=\SI{2}{\tera\electronvolt}$. This value is in agreement with lower bounds on singly-produced VLQs coupled to third generation quarks ~\cite{ATLAS:2025bzt, CMS:2024bni} and with the bounds from VLQ pair production~\cite{CMS:2024bni,ATLAS:2022hnn, ATLAS:2018ziw}. 

The interaction Lagrangians for the considered models are given for a general flavour structure in Ref.~\cite{deBlas:2017xtg} 
\begin{align}
    -\mathcal{L}_{\rm M.1}^{(\rm int)} =& \lsquare\lambda_U\rsquare_{rp} \bar{U}_R^r \tilde{H}^\dagger     q_L^p + \lsquare\lambda^u_{Q_1}\rsquare_{rp} \bar{Q}^r_{1L}\tilde{H}u^p_R + \lsquare\lambda^d_{Q_1}\rsquare_{rp} \bar{Q}_{1L}^rHd_R^p +\nonumber \\
        &+ \lsquare\lambda_{UQ_1}\rsquare_{rp} \bar{U}^r\tilde{H}^\dagger Q_1^p + \hc\,, 
\end{align}
\begin{equation}        
    -\mathcal{L}_{\rm M.2}^{(\rm int)} = \lsquare\lambda_U\rsquare_{rp} \bar{U}^r_R \tilde{H}^\dagger q^p_L + \lsquare\lambda_{Q_7}\rsquare_{rp}\bar{Q}^r_{7L} H u^p_R + \lsquare\lambda_{UQ_7}\rsquare_{rp} \bar{U}^rH^\dagger Q^p_7 + \hc \,, 
\end{equation}
\begin{align}    
    -\mathcal{L}^{(\rm int)}_{\rm M.3} =&  \lsquare\lambda^u_{Q_1}\rsquare_{rp}\bar{Q}^r_{1L}\tilde{H}\, u^p_{R} + \lsquare\lambda^d_{Q_1}\rsquare_{rp} \bar{Q}^r_{1L} H\, d^p_{R} +\frac{\lsquare\lambda_{T_1}\rsquare_{rp}}{2}\,\bar{T}_{1R}^{I\,r} H^\dagger \tau^I\, q^p_{L} + \nonumber \\ 
        &+\frac{\lsquare\lambda_{T_1 Q_1}\rsquare_{rp}}{2}\,\bar{T}_{1}^{I\, r} H^\dagger \tau^I\, Q^p_{1}+ \hc \,, 
\end{align}
\begin{align}        
    -\mathcal{L}^{(\rm int)}_{\rm M.4} =& \lsquare\lambda^u_{Q_1}\rsquare_{rp}\bar{Q}^r_{1L}\tilde{H}\, u^p_{R} + \lsquare\lambda^d_{Q_1}\rsquare_{rp} \bar{Q}^r_{1L} H\, d^p_{R} +\frac{\lsquare\lambda_{T_2}\rsquare_{rp}}{2}\,\bar{T}_{2R}^{I\,r} \tilde{H}^\dagger \tau^I\, q^p_{L} +\nonumber \\
        &+\frac{\lsquare\lambda_{T_2 Q_1}\rsquare_{rp}}{2}\,\bar{T}_{2}^{I\,r} \tilde{H}^\dagger \tau^I\, Q^p_{1}+ \hc \,,
\end{align}
\begin{equation}        
    -\mathcal{L}^{(\rm int)}_{\rm M.5} =  \lsquare\lambda_{Q_7}\rsquare_{rp} \bar{Q}^r_{7L} H u^p_R  +\frac{\lsquare\lambda_{T_2}\rsquare_{rp}}{2}\,\bar{T}_{2R}^{I\,r} \tilde{H}^\dagger \tau^I\, q^p_{L} + \frac{\lsquare\lambda_{T_2Q_7}\rsquare_{rp}}{2}\bar{T}_2^{I\,r} H^\dagger  \tau^I Q^p_7 + \hc\,.
\end{equation}
For these equations we follow the notation of Ref.~\cite{deBlas:2017xtg} and use $u_R^p$ and $d_R^p$ for right-handed quark singlets and $q_L^p$ for the left-handed quark doublets, being $p,r$ a flavour index. For our phenomenological study, we only consider third generation SM quarks, namely $p=r=3$.

For the tree-level expressions we refer to Ref.~\cite{deBlas:2017xtg}, whereas for the one-loop matching we use \texttt{SOLD}~\cite{Guedes:2023azv,Guedes:2024vuf} and \texttt{Matchete}~\cite{Fuentes-Martin:2022jrf}. 

\begin{figure}[t]
    \centering
    \begin{subfigure}[b]{0.46\textwidth}
        \centering
        \includegraphics[width = \textwidth]{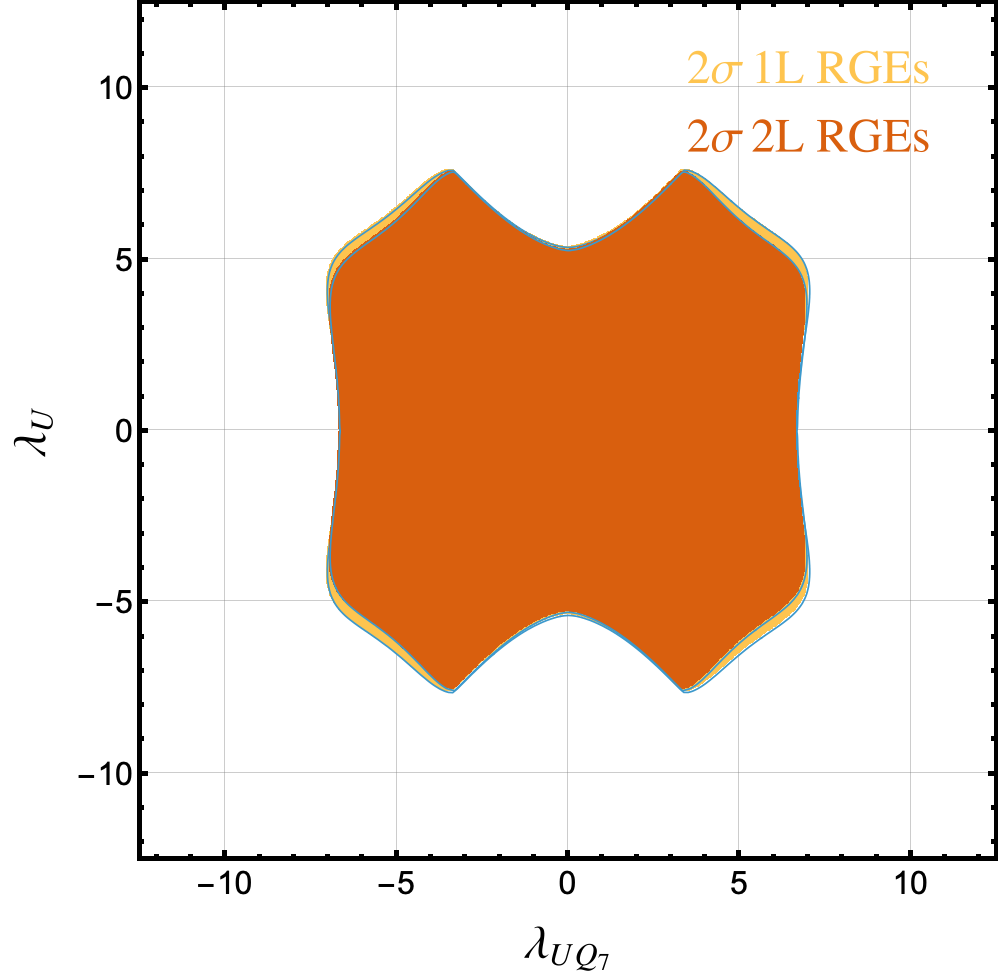}
        \caption{Model 2. }\label{fig:topdown_model2_v1}
    \end{subfigure}
    \hfill
    \begin{subfigure}[b]{0.46 \textwidth}
        \centering
        \includegraphics[width = \textwidth]{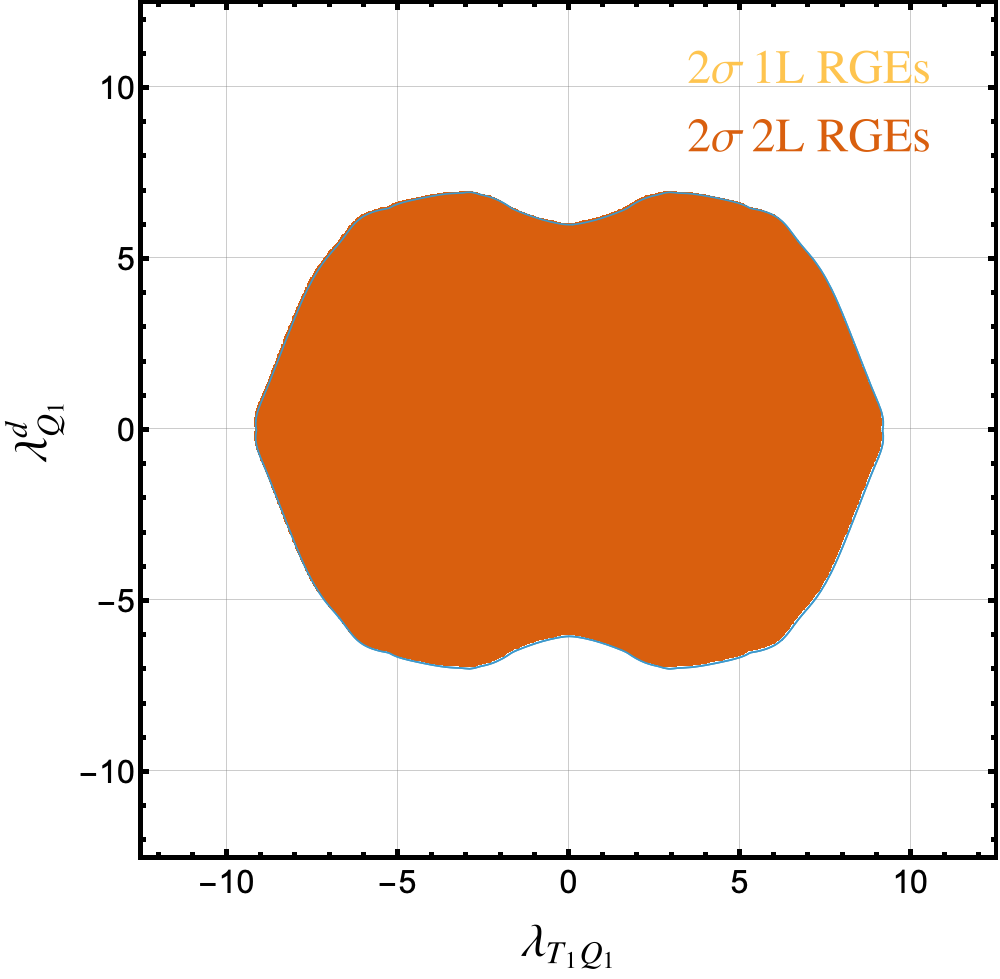}
        \caption{Model 3. }\label{fig:topdown_model3}
    \end{subfigure}\\
    \vspace*{1.1em}
    \begin{subfigure}[b]{0.46 \textwidth}
        \centering
        \includegraphics[width = \textwidth]{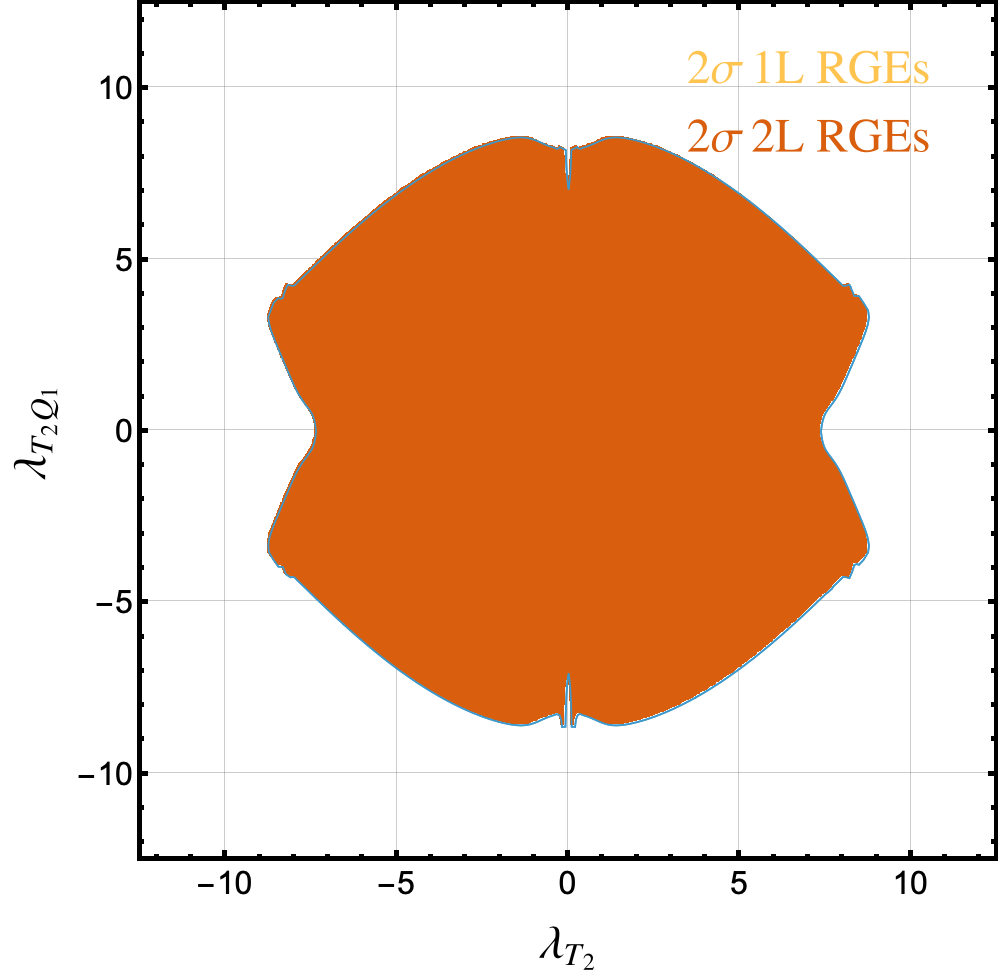}
        \caption{Model 4. }\label{fig:topdown_model4}
    \end{subfigure} 
    \hfill
    \begin{subfigure}[b]{0.46 \textwidth}
        \centering
        \includegraphics[width = \textwidth]{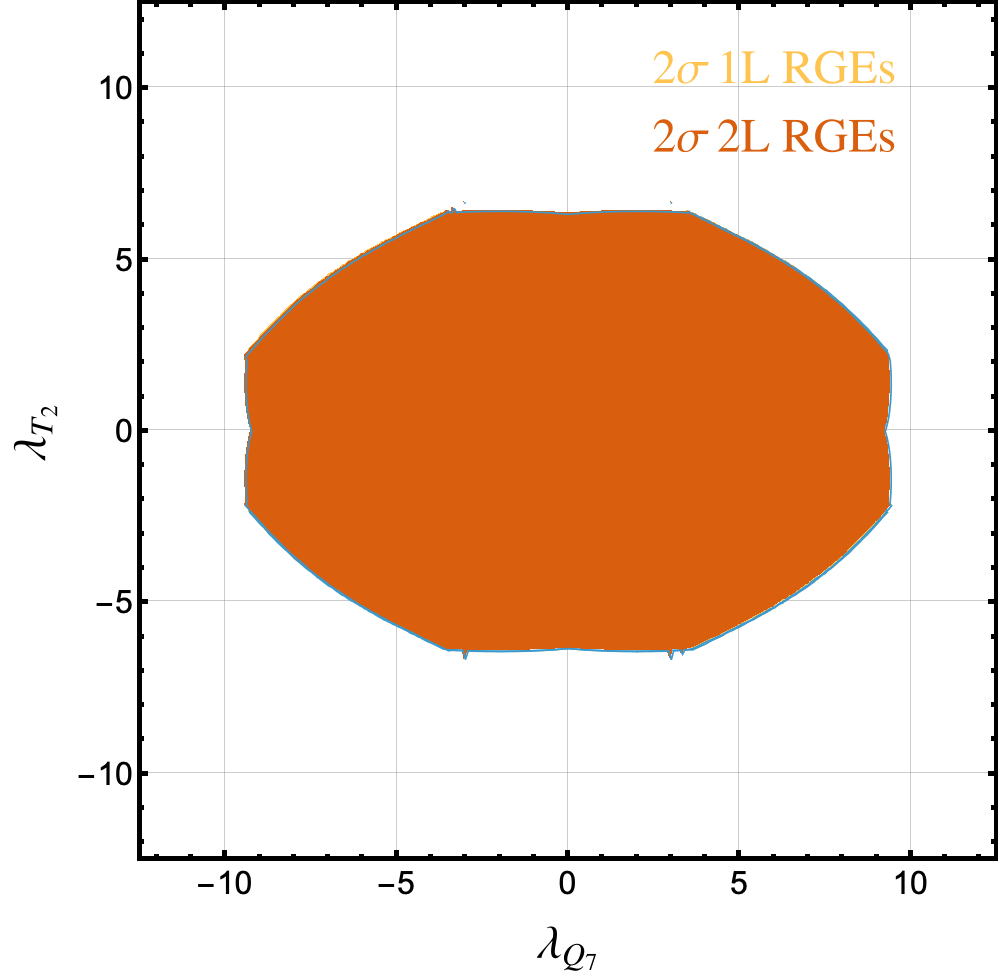}
        \caption{Model 5.}\label{fig:topdown_model5}
    \end{subfigure}
    \caption{Allowed regions at $2\sigma$ for pairs of couplings for the considered models. The orange (yellow) coloured areas correspond to allowed values derived accounting for two-loop (one-loop) running effects. \label{fig:topdown}}
\end{figure}

In all models and for all operators entering Eqs.~\eqref{eq:HiggsProductionCrossSection}-\eqref{eq:hllnunu} we match up to the one-loop level.

\begin{figure}
    \centering
     \begin{subfigure}[b]{0.46\textwidth}
        \centering
        \includegraphics[width = \textwidth]{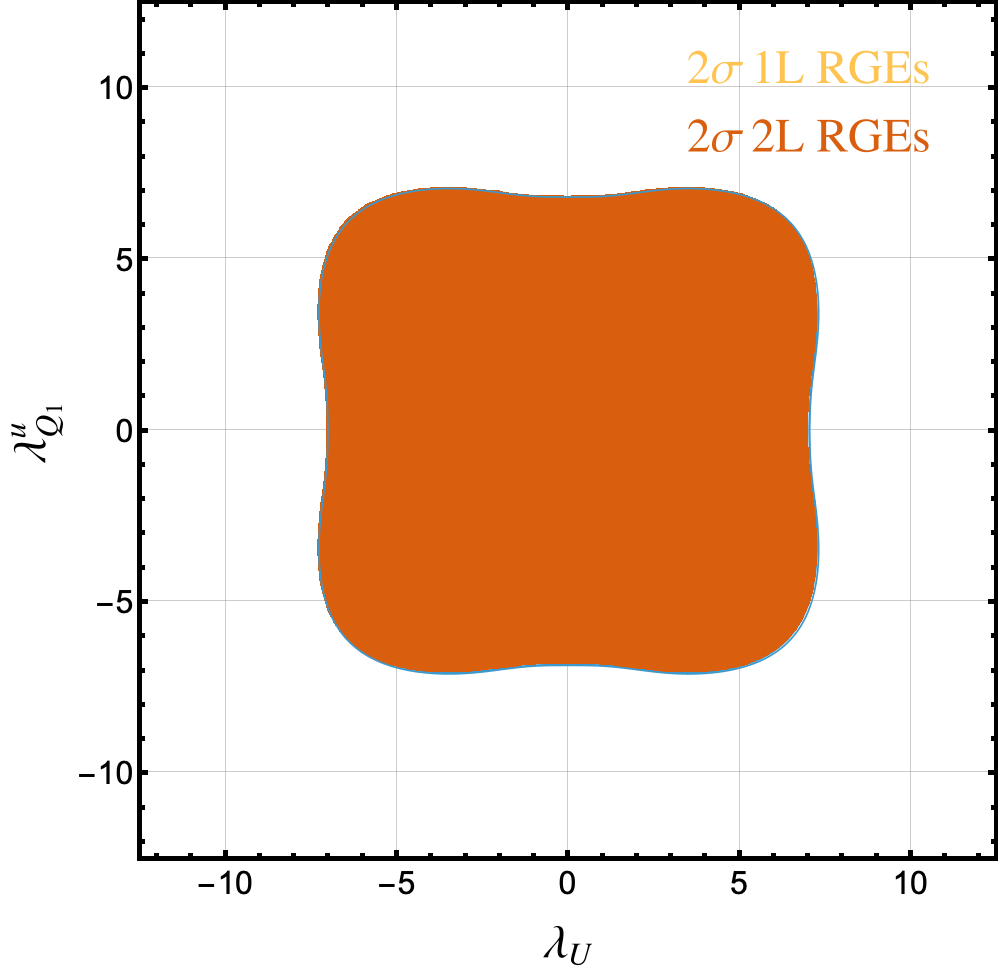} 
        \caption{Model 1 with complete one-loop matching }\label{fig:topdown_model1}
    \end{subfigure}
     \hfill
    \begin{subfigure}[b]{0.46 \textwidth}
        \centering
        \includegraphics[width = \textwidth]{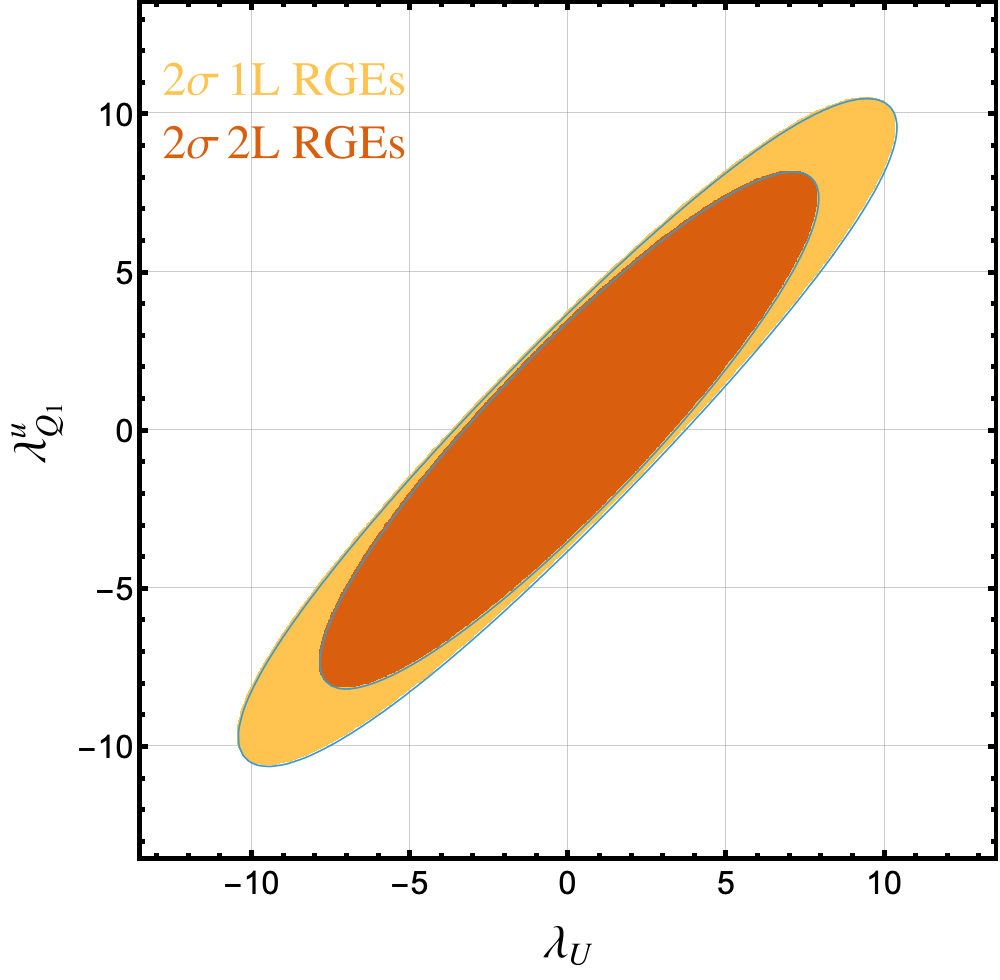}
        \caption{Model 1 using tree-level matching}\label{fig:topdown_model1_tl}
    \end{subfigure} 
    \caption{Two fits performed for Model 1, considering either the complete matching up to one loop for all the considered operators (left) or only the tree-level matching results (right). In both cases, the allowed parameter space at $2\sigma$ is plotted in yellow (orange) when the one-loop (two-loop) running effects are included. }
    \label{fig:topdown_comparison}
\end{figure}

For each model the $\chi^2$ is a function of either three or four couplings in the UV model. In Fig.~\ref{fig:topdown} and Fig.~\ref{fig:topdown_model1} we present our results as two-dimensional CL regions, where the remaining couplings were profiled over.
As can be inferred in Figs.~\ref{fig:topdown}--\ref{fig:topdown_model1}, the inclusion of two-loop running in the RGEs makes only a minor difference. While the model at tree-level matches only into operators of class 7 and the Yukawa-type operators, at one-loop level it matches into the class 3 operators and $\op{HG}{}$, $\op{HW}{}$, $\op{HB}{}$ and $\op{HWB}{}$, which enter with large coefficients into the signal strength. This reduces the impact of the two-loop running effects considered here. A greater impact is observed when only the tree-level matching results are considered. In order to showcase this aspect and to explain why the effect in the concrete models is so much smaller than the one shown in Fig.~\ref{fig:results_oneparameter_1Lvs2L}, we compare in Fig.~\ref{fig:topdown_comparison} the complete fit performed for model 1 with a second fit in which only the tree-level generated operators are considered. 
We note that of course performing the matching at one-loop level is the consistent way to proceed, hence Fig.~\ref{fig:topdown_model1_tl}\footnote{For simplicity, the plot in this case is obtained by setting the couplings not shown to their best-fit value.} is for illustrative purpose only.
Before moving on to our conclusions, we would like to make two comments on the plots  shown in this section. First of all, our results should not be taken as competitive bounds on the allowed parameter space for these UV models but as a way to test the impact of running effects. Secondly, we provide a brief motivation for the peculiar shapes in Figs~\ref{fig:topdown_model3}--\ref{fig:topdown_model4}. Noticing the hyperbolic features of the borders and assuming the effect to be due to the less suppressed tree-level generated coefficients, we identify the Yukawa-like operators $\coeff{tH}{}$ and $\coeff{bH}{}$ as the only ones whose tree-level matching involves the product of three UV couplings. When one is fixed, the remaining give rise to an approximate hyperbola. However, while for the other models only $\coeff{tH}{}$ has such contribution, for the two models involving the doublet $Q_1$ and a triplet both Yukawa-like operators involve this term. To further explain this point, below we compare the tree-level matching for models 1 and 3. For model 1 one has
\begin{align}
    \coeff{tH}{\rm M.1} &\overset{\rm T.L.}{=} -\frac{1}{2\Lambda^2} \left(-\yuk{t} \vert\lambda_U\vert^2  -\yuk{t} \vert\lambda^u_{Q_1}\vert^2 + 2  \lambda^u_{Q_1} \lambda_U \lambda_{UQ_1}\right)\,,\\
    \coeff{bH}{\rm M.1} & \overset{\rm T.L.}{= } \frac{1}{2 \Lambda^2}\yuk{b} \vert\lambda^d_{Q_1}\vert^2\,.
\end{align}
Instead, for model 3, Ref.~\cite{deBlas:2017xtg} provides
\begin{align}
    \coeff{tH}{\rm M.3} & \overset{\rm T.L.}{= }  - \frac{1}{4\Lambda^2} \left(-2 \yuk{t} \vert\lambda^u_{Q_1}\vert^2  -\yuk{t}   \vert\lambda_{T_1}\vert^2 + 2 \lambda_{T_1Q_1} \lambda^u_{Q_1} \lambda_{T_1}\right)\,, \\
    \coeff{bH}{\rm M.3}&\overset{\rm T.L.}{=} \frac{1}{8\Lambda^2} \left(4 \yuk{b} \vert\lambda^d_{Q_1}\vert^2 +  \yuk{b}  \vert \lambda_{T_1}\vert^2 - 2 \lambda^d_{Q_1} \lambda_{T_1Q_1} \lambda_{T_1} \right)\,.
\end{align}
To test this point, we have studied the two models setting $\coeff{fH}{}=0$ for $f=t,b$ and found more regular parameter spaces. However, these are unphysical and are therefore not included.

\section{Conclusions} \label{sec:conclusions}
We have analyzed the two-loop RGE of the effective Higgs-gluon coupling in the SMEFT, parametrized by the operator $\op{HG}{}$. 
Several single and double Higgs production channels are directly sensitive to this coupling, that enters already at tree level whereas the SM signal arises at one loop. 

At one loop, $\coeff{HG}{}$ renormalizes itself and also mixes with the chromomagnetic operator. Adopting a loop counting in weakly coupled, renormalizable UV completions of the SM \cite{Arzt:1994gp,Buchalla:2022vjp}, these operators  are expected to arise only at the one-loop level, pushing their RGE effects at two-loop order. Consequently, operators that can be generated at tree level and mix with $\coeff{HG}{}$ at two loop must also be consistently included under this counting.
Our computation therefore extends the previous efforts for the computation of the two-loop contributions of the potentially tree-level generated four-top~\cite{DiNoi:2023ygk} and Yukawa-like operators~\cite{DiNoi:2024ajj}, presenting for the first time the two-loop contributions of the operators with schematic structure $D^2 H^4$ and $D\, H^2 \,\psi^2$.

We addressed the phenomenological impact of the two-loop running in a fit to inclusive Higgs data. Indeed, we could find that bounds on the class 7 operators  shrink  due to the two-loop running effects, since their contribution is proportional to the top Yukawa coupling. In concrete models, such as the ones with vector-like quarks presented in our phenomenological study, this effect is reduced, as many more operators arise when matching up to one-loop order. For instance, $\op{HG}{}$ is generated at one-loop level in this type of models. 

Nevertheless, two-loop running effects could have an impact on global fits, once completely available. This applies in particular for operators so far loosely constrained. For instance, in the case of four-top operators, it was shown that two-loop corrections to electroweak precision observables \cite{Haisch:2024wnw} can, in some cases, limit the currently allowed range \cite{DiNoi:2025uhu}.

To summarize, our results complete an important piece of the two-loop RGE framework for Higgs–gluon interactions, setting the stage for consistent and precise SMEFT analyses of Higgs data with potential phenomenological relevance.

\section*{Note added}
During the completion this work, we became aware of \cite{Javierinprep}. Our results are in agreement.

\section*{Acknowledgements}
 We thank Javier Fuentes-Mart\'in for cross-checking our results with \cite{Javierinprep} and the discussions that made us realise that we need to clearly state the definition of the SM Lagrangian that we adopt.
 BE  would like to thank Jonathan Ronca for several useful discussions on numerical integration and Alejo~N. Rossia for comments on marginalisation and profiling. Also, we would like to acknowledge the Mainz Institute for Theoretical Physics (MITP) of the Cluster of Excellence PRISMA+ (Project ID 390831469) for its hospitality and its partial support during part of the completion of this work.
This work received funding by the INFN Iniziativa Specifica APINE and by the University of Padua under the 2023 STARS Grants@Unipd programme (Acronym and title of the project: HiggsPairs – Precise Theoretical Predictions for Higgs pair production at the LHC). This work was also partially supported by the Italian MUR Departments of Excellence grant 2023-2027 “Quantum Frontiers”. The research of SDN was supported by the Deutsche Forschungsgemeinschaft (DFG, German Research Foundation) under grant 396021762 - TRR 257 “Particle
Physics Phenomenology after the Higgs Discovery”. We acknowledge support by the COST Action COMETA CA22130. \\
CloudVeneto is acknowledged for the use of computing and storage facilities. The Feynman diagrams in this work were made using \texttt{TikZ-Feynman} \cite{Ellis:2016jkw}.
\appendix

\section{Feynman Rules} \label{app:feynmanrules}
In this Appendix we present the Feynman rules associated with the effective operators defined in Eqs.~\eqref{eq:psi2phi2Dops_phiqL}-\eqref{eq:psi2phi2Dops_phiud}. $m_j$ ($A_j$) identifies an index in the fundamental (adjoint) representation of $\mathrm{SU(3)}_{\rm C}$ whereas $i_j$ denotes an index in the fundamental representation of $\mathrm{SU(2)}_{\rm L}$. The numbers are associated to the particles in the figure on the left. 
\begin{subequations}
\begin{align}
\begin{tikzpicture}[baseline=(hhgg)]
            \begin{feynman}
                \vertex  (h1)  {$H^{3,*}$};
                \vertex  (hhgg) [dot,magenta,below left= of h1] {};
                \vertex  (h2) [below right = of hhgg] {$H^4$}; 
                \vertex (g1) [above left =  of hhgg] {$g^1$};
                \vertex (g2) [below left =  of hhgg] {$g^2$};
                \diagram* {
                (g1) -- [gluon] (hhgg) -- [gluon] (g2),
                (h1) -- [scalar] (hhgg) -- [scalar] (h2),
                };
            \end{feynman}
             \node[shape=regular polygon,regular polygon sides=4, aspect=1,fill=white, draw=black,scale = .5,opacity=1.] at (hhgg) {};
        \end{tikzpicture} &=  + 4 i \; \coeff{H G}{}  \delta_{i_3 i_4} \delta_{A_1 A_2} \left(p_1^{\mu_2}p_2^{\mu_1} - g^{\mu_1 \mu_2} p_1 \cdot p_2 \right).\\
\begin{tikzpicture}[baseline = (center)]
            \begin{feynman}
                \vertex (center) [black, square dot, scale=\sizesqdot] at (0,0) {};
                \vertex[above right = of center] (phi1) {\(H^1\)};
                \vertex[below right = of center] (phi2) {\(H^{2,\,*}\)};
                \vertex[above left = of center] (qL) {\(Q_L^4\)};
                \vertex[below left = of center] (qbarL) {\({Q}_L^3\)};
                \diagram*{
                    (qL) -- [fermion, line width=\lwL] (center) -- [fermion, line width=\lwL] (qbarL);
                    (phi1) -- [scalar] (center) -- [scalar] (phi2); 
                };
            \end{feynman}
        \end{tikzpicture} &= + i\;  \left(\coeff{H q}{(1)} \delta_{i_1i_2} \delta_{i_3i_4}  +  \coeff{H q}{(3)} \tau^I_{i_2i_1} \tau^I_{i_4i_3}  \right) \delta_{m_3 m_4} \lround \slashed{p}_1 - \slashed{p}_2 \rround \mathbb{P}_{\rm L} , \\
\begin{tikzpicture}[baseline = (center)]
            \begin{feynman}
                \vertex (center) [black, square dot, scale=\sizesqdot]  at (0,0) {};
                \vertex[above right = of center] (phi1) {\(H^1\)};
                \vertex[below right = of center] (phi2) {\(H^{2,\,*}\)};
                \vertex[above left = of center] (bR) {\(\psi_R^3\)};
                \vertex[below left = of center] (bbarR) {\({\psi}_R^4\)};
                \diagram*{
                    (bR) -- [fermion, line width=\lwR] (center) -- [fermion, line width=\lwR] (bbarR);
                    (phi1) -- [scalar] (center) -- [scalar] (phi2); 
                };
            \end{feynman}
        \end{tikzpicture} &= + i\; \coeff{H \psi}{} \delta_{i_1i_2} \delta_{m_3m_4} \lround \slashed{p}_1 - \slashed{p}_2 \rround \mathbb{P}_{\rm R} ,  \qquad \psi = t,b \; ,\\  
\begin{tikzpicture}[baseline = (center)]
            \begin{feynman}
                \vertex (center) [black, square dot, scale=\sizesqdot] at (0,0) {};
                \vertex[above right = of center] (phi1) {\(H^1\)};
                \vertex[below right = of center] (phi2) {\(H^{2}\)};
                \vertex[above left = of center] (dR) {\(d_R^4\)};
                \vertex[below left = of center] (ubarR) {\(u_R^3\)};
                \diagram*{
                    (dR) -- [fermion, line width=\lwR] (center) -- [fermion, line width=\lwR] (ubarR);
                    (phi1) -- [scalar] (center) -- [scalar] (phi2); 
                };
            \end{feynman}
        \end{tikzpicture} &= + i\;  \coeff{Htb}{} \delta_{m_3m_4} \epsilon_{i_1i_2}\lround \slashed{p_1}- \slashed{p_2} \rround \mathbb{P}_{\rm R}, \\
\begin{tikzpicture}[baseline = (center)]
            \begin{feynman}
                \vertex (center) [black, square dot, scale=\sizesqdot] at (0,0) {};
                \vertex[above right = of center] (phi1) {\(H^1\)};
                \vertex[below right = of center] (phi2) {\(H^{2}\)};
                \vertex[above left = of center] (phi4) {\( H^{4,*}\)};
                \vertex[below left = of center] (phi3) {\(H^{3, *}\)};
                \diagram*{
                    (phi4) -- [scalar] (center) -- [scalar] (phi3);
                    (phi1) -- [scalar] (center) -- [scalar] (phi2); 
                };
            \end{feynman}
        \end{tikzpicture} &=  \text{See Eq.~\eqref{eq:FeynmanRules_D2phi4}.}
\end{align}
\end{subequations}
Given their length, we report here the Feynman rules for the last diagrams above:
\begin{equation}
     -i \left[ \delta_{i_1i_4}\delta_{i_2i_3}\left( \coeff{HD}{}\left( p_1\cdot p_3 + p_2\cdot p_4 \right) + \coeff{H\Box}{} \left( \sum_i p_i^2 + 2p_1\cdot p_4 + 2 p_2\cdot p_3  \right) \right) + \left(4\leftrightarrow 3\right) \right].
\label{eq:FeynmanRules_D2phi4}\end{equation}

For the sake of completeness we present the Feynman rules of the relevant SM interactions. We remind the reader that the for the Yukawa sector of the SM we adopt the same notation of Refs.~\cite{rge1, rge2, rge3}.
\begin{subequations}
\begin{alignat}{4}\label{FR:yu}
 \begin{tikzpicture}[baseline=(gqq)]
            \begin{feynman}[small]
                \vertex  (g)  {$g^1$};
                \vertex  (gqq) [dot, scale=\sizedot,right= of g] {};
                \vertex (q1) [above right =  of gqq] {$Q_L^2$};
                \vertex (q2) [below right =  of gqq] {$Q_L^3$};
                \diagram* {
                    (g)  -- [gluon] (gqq),
                    (q1) -- [fermion, line width=\lwL] (gqq) -- [fermion, line width=\lwL] (q2)
                };
            \end{feynman}
        \end{tikzpicture} &=i g_s \, \gamma^{\mu_1 } \PL T^{A_1}_{m_3m_2} \delta_{i_3 i_2},\quad
        \begin{tikzpicture}[baseline=(gqq)]
            \begin{feynman}[small]
                \vertex  (g)  {$g^1$};
                \vertex  (gqq) [dot, scale=\sizedot,right= of g] {};
                \vertex (q1) [above right =  of gqq] {$t_R^2/b_R^2$};
                \vertex (q2) [below right =  of gqq] {$t_R^3/b_R^3$};;
                \diagram* {
                    (g)  -- [gluon] (gqq),
                    (q1) -- [fermion, line width=\lwR] (gqq) -- [fermion, line width=\lwR] (q2)
                };
            \end{feynman}
        \end{tikzpicture} =i g_s \, \gamma^{\mu_1 } \PR T^{A_1}_{m_3m_2},
        \\        
 \begin{tikzpicture}[baseline=(hqq)]
            \begin{feynman}[small]
                \vertex  (h)  {$H^{1}$};
                \vertex  (hqq) [dot, scale=\sizedot,right= of h] {};
                \vertex (q1) [above right =  of hqq] {$Q^2_L$};
                \vertex (q2) [below right =  of hqq] {$t^3_R$};
                \diagram* {
                    (h)  -- [scalar] (hqq),
                    (q1) -- [fermion, line width=\lwL] (hqq) -- [fermion, line width=\lwR] (q2)
                };
            \end{feynman}
        \end{tikzpicture} &= + i \yuk{t} \, \PR \delta_{m_3 m_2} \varepsilon_{i_1 i_2}  , \qquad \, \begin{tikzpicture}[baseline=(hqq)]
            \begin{feynman}[small]
                \vertex  (h)  {$H^{1,*}$};
                \vertex  (hqq) [dot, scale=\sizedot,right= of h] {};
                \vertex (q1) [above right =  of hqq] {$t_R^2$};
                \vertex (q2) [below right =  of hqq] {$Q_L^3$};
                \diagram* {
                    (h)  -- [scalar] (hqq),
                    (q1) -- [fermion, line width=\lwR] (hqq) -- [fermion, line width=\lwL] (q2)
                };
            \end{feynman}
        \end{tikzpicture} =  -i \yuk{t}^* \PL \delta_{m_3 m_2}   \varepsilon_{i_3 i_1},
        \\
  \begin{tikzpicture}[baseline=(hqq)]
            \begin{feynman}[small]
                \vertex  (h)  {$H^{1,*}$};
                \vertex  (hqq) [dot, scale=\sizedot,right= of h] {};
                \vertex (q1) [above right =  of hqq] {$Q^2_L$};
                \vertex (q2) [below right =  of hqq] {$b^3_R$};
                \diagram* {
                    (h)  -- [scalar] (hqq),
                    (q1) -- [fermion, line width=\lwL] (hqq) -- [fermion, line width=\lwR] (q2)
                };
            \end{feynman}
        \end{tikzpicture} &=  -i \yuk{b} \,\PR \delta_{m_3 m_2} \delta_{i_1 i_2}  , \qquad  \; \begin{tikzpicture}[baseline=(hqq)]
            \begin{feynman}[small]
                \vertex  (h)  {$H^{1}$};
                \vertex  (hqq) [dot, scale=\sizedot,right= of h] {};
                \vertex (q1) [above right =  of hqq] {$b_R^2$};
                \vertex (q2) [below right =  of hqq] {$Q_L^3$};
                \diagram* {
                    (h)  -- [scalar] (hqq),
                    (q1) -- [fermion, line width=\lwR] (hqq) -- [fermion, line width=\lwL] (q2)
                };
            \end{feynman}
        \end{tikzpicture} =  -i  \yuk{b}^* \PL \delta_{m_3 m_2} \delta_{i_3 i_1} .
\end{alignat}
\end{subequations}

\bibliographystyle{JHEP.bst}
\bibliography{bibliography.bib}

\end{document}